\documentclass[superscriptaddress, reprint, twocolumn%
reprint,
amsmath,amssymb,
aps,
]{revtex4-2}

\usepackage{graphicx}
\usepackage{dcolumn}
\usepackage{booktabs}
\usepackage{tabularx}
\usepackage[mathlines]{lineno}

\usepackage{placeins} 
\usepackage{bm}
\usepackage{fixltx2e}

\usepackage{color}

\begin{document}
	
	\preprint{APS/123-QED}
	
	\title{Effect of Interlayer Stacking on the Electronic Properties of 1\emph{T}-TaS$_{2}$}
	
	\author{Nelson Hua}
	\email{nelson.hua@psi.ch}
	\affiliation{PSI Center for Photon Science, Paul Scherrer Institute, 5232 Villigen PSI, Switzerland}%
 	\affiliation{Institute for Quantum Electronics, ETH Zurich, 8093 Zurich, Switzerland}
    \author{Francesco Petocchi}
	\affiliation{Department of Quantum Matter Physics, University of Geneva, 1211 Geneva, Switzerland}
	\author{Henry G. Bell}
	\affiliation{PSI Center for Photon Science, Paul Scherrer Institute, 5232 Villigen PSI, Switzerland}%
	\affiliation{Laboratory for Solid State Physics and Quantum Center, ETH Zurich, 8093 Zurich, Switzerland}
    \author{Gabriel~Aeppli}
    \affiliation{PSI Center for Photon Science, Paul Scherrer Institute, 5232 Villigen PSI, Switzerland}
    \affiliation{Laboratory for Solid State Physics and Quantum Center, ETH Zurich, 8093 Zurich, Switzerland}
    \affiliation{Institute of Physics, EPF Lausanne, 1015 Lausanne, Switzerland}  

	\author{Philipp Werner}
    \email{philipp.werner@unifr.ch}%
	\affiliation{Department of Physics, University of Fribourg, 1700 Fribourg, Switzerland}
 	\author{Simon Gerber}
    \email{simon.gerber@psi.ch}%
	\affiliation{PSI Center for Photon Science, Paul Scherrer Institute, 5232 Villigen PSI, Switzerland}
	\date{\today}
	
	\begin{abstract}
    {
		Controlled stacking of van der Waals materials is a powerful tool for exploring the physics of quantum condensed matter. Given the small binding between layers, exploitation for engineering will require a breakthrough in stacking methodology, or an ability to take advantage of thicker defective stacks. Here we describe computational groundwork for the latter, using---on account of its promise for cold memory applications---1\textit{T}-TaS$_2$ as a model system. Comparing recursive Hendricks-Teller calculations and Monte Carlo simulations to published X-ray diffraction data, we obtain the key parameters describing the random stacking in mesoscopic flakes. These then regulate the electronic structures via specification of the random stacks in dynamical mean-field theory simulations. Hubbard repulsion induces strongly correlated metallic, band and Mott insulating layers, providing compelling evidence that electronic properties follow from the coexistence of more than the metallic and insulating planes associated by ordinary band theory. 

        }
	\end{abstract}
	
	\maketitle

Strong electronic correlations in single layers of van der Waals (vdW) materials can lead to intriguing properties not present in bulk counterparts. For instance, topologically protected edge states arising from the quantum spin Hall effect are predicted for a class of two-dimensional transition-metal dicalcogenides~\cite{Qian2014}, and have recently been demonstrated in a monolayer of WTe$_{2}$~\cite{Fei2017}. Isolating such systems is possible due to the weak vdW interaction between layers that allows for exfoliation, providing stacking flexibility to construct novel heterostructures not feasible with other fabrication techniques \cite{Geim2013}. However, even though the interlayer coupling strength is relatively weak, a particular stacking structure can still influence the emergent macroscopic properties. For example, the difference between AA and AB stacking in bilayer graphene significantly alters the range of bandgap tunability \cite{Ould2017,Liang2020}. Furthermore, the ability to twist bilayer graphene into a Moiré pattern ---where a magic angle leads to a flat band---opens pathways to unconventional superconductivity and correlated insulators \cite{Bistritzer2011, Cao2018}. Resolving the stacking structures and understanding their influence on the electronic properties is therefore a crucial step towards precise control and efficient heterostructure design~\cite{Liu2016, Wang2023_1T-TaSe2}.

A particular system for which there is both scientific and emerging technological interest related to stacking is the classic vdW material 1\emph{T}-TaS$_{2}$, long known for its commensurate charge-density wave (CCDW) phase. Here, the discovery of a hidden, nonthermal metallic state~\cite{Stojchevska2014} that can be switched either optically or electrically has provided a new concept for memory devices~\cite{Wang2012, Jariwala2014, Liu2016}, motivating considerable research that recently showed the switching to be a bulk effect associated with interlayer restacking  \cite{Stahl2020, Burri2024}. Further progress in both the science and engineering aspects requires quantitative modeling of the stacking and its consequences for the electronic structure of the material, which are the topics of the present paper.  

The various electronic phases of 1\emph{T}-TaS$_{2}$, appearing as a function of temperature, are primarily characterized by the degree of \textit{in-plane} commensurability of so-called polaron stars---motifs where \mbox{12 Ta atoms} contract towards a central Ta~atom (see Fig.~1). This has been confirmed through various experimental probes, such as transmission electron microscopy~(TEM)~\cite{Tsen2015, Wang2023}, scanning tunneling microscopy (STM) \cite{Ma2016,Gerasimenko2019, Ravnik2021} and X-ray diffraction~(XRD)~\cite{Tanda1984, Nakanishi1984, Stahl2020, Burri2024}. 

The \textit{interlayer} coordination of the polaron stars can be described by \emph{T$_{a}$}~(dimer/bilayers), \emph{T$_{b}$}, and \emph{T$_{c}$} stacking (see Fig.~1). The exact distribution of stacking configurations in bulk 1\emph{T}-TaS$_{2}$ remains undetermined  as it is hidden to experimental techniques with limited subsurface sensitivity. TEM shows all three stacking configurations at the surface \cite{Wang2023} while STM reveals predominantly \emph{T$_{a}$} and \emph{T$_{c}$} stacking with minority regions of~\emph{T$_{b}$}~\cite{Butler2020}. Since the exfoliation process is inherently invasive and the surface properties may deviate from the bulk structure, it is questionable to extrapolate these findings beyond the first few layers. Similarly, angle-resolved photoemission spectroscopy (ARPES) probes primarily the surface for vacuum-ultraviolet light, and has thus far yielded conflicting results for the electronic ground state that is either band or Mott insulating~\mbox{\cite{Perfetti2006, Hellmann2010, Hellmann2012, Ritschel2015, Wang2020, Yang2024, Wang2024}}. When investigating coherent dynamics with time-resolved techniques~\mbox{\cite{Sugai1981, Toda2004, Ravnik2017, Tanimura2018}}, disagreements in the assignment of phonon modes persist. As demonstrated with density functional theory (DFT) calculations~\cite{Albertini2016}, the spectrum of phonon modes depends heavily on the exact stacking configuration. Several studies have also postulated buried monolayers within the bulk that host quantum spin liquids~(QSL)~\cite{Tsen2015, Klanjvsek2017, Law2017, Kratochvilova2017, Manas-Valero2021}. In contrast, other reports assumed predominant dimerization in the bulk \cite{Ritschel2015, Ritschel2018, Lee2019, Stahl2020, Butler2020}, precluding a QSL phase. 

\begin{figure}[!t]
	\includegraphics[width=\columnwidth]{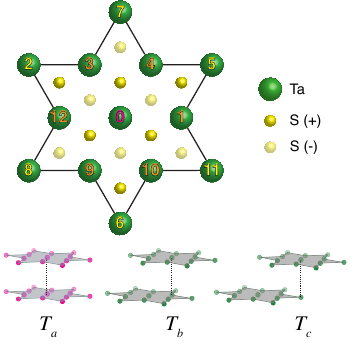}
    
	\caption{\label{Fig1a}
	Polaron star with Ta atoms indexed from 0 to 12. For \emph{T$_{a}$}, \emph{T$_{b}$} and \emph{T$_{c}$} configurations, the stacking of layers is referenced to the central Ta atom~0 (pink). 
    The staggered S~atoms above ($+$) and below ($-$) the Ta~plane break the six-fold symmetry: \emph{T$_{b}$} can be subdivided into \emph{T$_{b1}$} and \emph{T$_{b2}$} corresponding to stacking above Ta atoms \{1,~3,~9\} and \{4,~10,~12\} (orange), respectively. The same applies for \emph{T$_{c1}$} and \emph{T$_{c2}$} stacking above Ta atoms \{2,~5,~6\} and \{7,~8,~11\} (yellow), respectively.}
\end{figure}

Notwithstanding their variable outcomes, the studies mentioned above all conclude that the interlayer configuration mediates the emergent electronic properties of \mbox{1\emph{T}-TaS$_{2}$}, underscoring the urgency to specify the stacking structure. The obvious method to determine the interlayer stacking order is XRD, due to its high penetration depth and resolution. However, despite several XRD experiments characterizing the Bragg peaks of the CCDW phase, a consensus has not been reached. Early studies \cite{Tanda1984, Nakanishi1984} carefully mapped out the peak positions associated with the various CDW phases of \mbox{1\emph{T}-TaS$_{2}$}. A disordered stacking was held accountable for the broad out-of-plane diffraction signals and half-ordered (dimer) peaks that are signatures of dimerization in the CCDW phase. But the exact stacking structure remained unspecified---though, it was hypothesized that the CCDW phase is characterized by a distribution of \emph{T$_{a}$}, \emph{T$_{b}$} and \emph{T$_{c}$} stacking. In addition to the peak broadening indicative of disorder, CDW peaks at $l = \frac{1}{5}$ reciprocal lattice units~(r.l.u.) \cite{Tanda1984, Nakanishi1984, Laulhe2015, Stahl2020} imply a periodicity of five layers that seems incompatible with the three-fold CDW symmetry. Surprisingly, this unusual value has not been scrutinized and, to our best knowledge, no structure factor simulations are available to discern the exact stacking structure.

For the science and engineering of disordered systems, the key variables are the probabilities of particular configurations, which can then be used as inputs to model physical properties.  Therefore, we start with   Monte Carlo (MC) simulations and a recursive variation of the \mbox{Hendricks-Teller~(HT)} method~\cite{Hendricks1942,Treacy1991} with structure factor calculations to reveal the probabilities underpinning the random stacking of 1\emph{T}-TaS$_{2}$ seen 
experimentally. 
After demonstrating the equivalency of the numerical MC and analytical HT approaches, 
MC is used to generate multi-layer configurations that reproduce the experimentally-observed out-of-plane scattering pattern. Dynamical mean-field theory~(DMFT) calculations of these configurations then provide access to the layer-resolved spectral functions of the bulk material, and allow us to address open questions about the emergent behaviors, such as the controversy of whether the insulating low-temperature ground state is of Mott or band insulating nature. An advantage of the DMFT approach over simple DFT calculations is that it considers electron correlations that we identify as drivers for the formation of Mott insulating monolayers in the low-temperature ground state of 1\emph{T}-TaS$_{2}$.

XRD patterns from periodic structures can be calculated from the structure factor given by
\begin{equation} \label{eq1}
F(\vec{Q}) =  \sum_{i = 1}^M \sum_{n = 1}^N f_{i,n}(Q) e^{-i\vec{Q} \cdot \vec{R}_{i,n}},
\end{equation}
which is the sum of the atomic form factor $f$ of all atoms and their position in the unit cell $\vec{R}$. The scattered intensity is then obtained by $I(\vec{Q}) = |F(\vec{Q})|^{2}$. Here the subscript $i$ denotes the in-plane coordinate while $n$ is the out-of-plane position of the atoms. Assuming no in-plane disorder, we can define the structure factor for a monolayer $F_\text{planar} (\vec{Q})$ by
\begin{equation} \label{eq2}
\begin{gathered}
    F_\text{planar} (\vec{Q}) = \sum_{i = 1}^M f_i(Q)e^{-i\vec{Q} \cdot \vec{r}_i} \\   
    F(\vec{Q}) = F_\text{planar}(\vec{Q}) \sum_{n=1}^N e^{-i\vec{Q} \cdot \Delta \vec{r}_n},
\end{gathered}
\end{equation}
where $\vec{R}_{i,n} = \vec{r}_i + \Delta \vec{r}_n$ is decomposed into its in- and out-of-plane components. 
This simplifies the calculation of layered systems, where $F_\text{planar} (\vec{Q})$ needs to be explicitly calculated only once, and the three-dimensional stacking structure is constructed by multiplying the sum of exponentials by the offset $\Delta \vec{r}_n$ of subsequent layers. 

In the case of \mbox{1\emph{T}-TaS$_{2}$}, there are 13 possible offsets corresponding to the \emph{T$_{a}$}, \emph{T$_{b}$}, and \emph{T$_{c}$} stacking structures. With this formalism, we can exhaust all possible stacking distributions by varying the probabilities that follow $p_{a} + p_{b} + p_{c} = 1$. Due to the staggered S atoms that break the six-fold symmetry, $p_{b}$ and $p_{c}$ can be further decomposed into $p_{b1,b2}$ and $p_{c1,c2}$. Such simulations can be done numerically with MC methods (see Supplementary Material) or analytically with a recursive variation of the HT method suited for modeling disorder in one dimension \cite{Hendricks1942,Treacy1991}. For a system with \emph{N}~layers, the scattered intensity at a given $\vec{Q}$ position is described by
\begin{equation} \label{Hendriks-Teller Result}
    I(\vec{Q})/N = \vec{G}^* \cdot \vec{\psi} +\vec{G} \cdot \vec{\psi}^* - \vec{G}^* \cdot \vec{F},    
\end{equation}
where the vector $G_i = g_i F_i$ and 
$$\vec{\psi} = \frac{(I-T)^{-1}}{N}\left[(N+1)I - (I - T)^{-1}(I-T^{N+1})\right] \vec{F}.$$
\emph{I} is the identity matrix. The complex transition  matrix $T_{ij}(\vec{Q}) = A_{ij} e^{i\vec{Q} \cdot \Delta \vec{r}_{ij}}$ is determined by the stacking probabilities for layers of type \emph{i} and \emph{j}. \emph{g$_{i}$} is a normalization factor to ensure the stacking probabilities sum to unity. 

\begin{figure}[!t]
	\includegraphics[width=1\columnwidth]{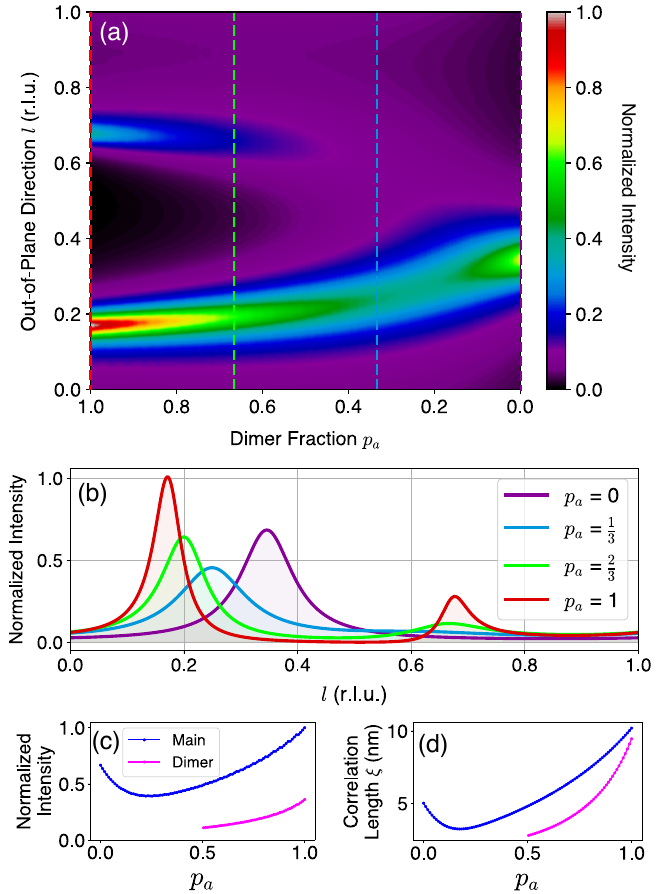}
    
	\caption{\label{Fig1b}
(a) Simulated XRD intensity as a function of the out-of-plane reciprocal lattice direction \emph{l} and the fraction of dimerized layers $p_a$ for a system with complete \emph{T$_{c}$} stacking. A half-ordered (dimer) peak appears when $p_a\gtrapprox$ 0.5. (b)~Selected line cuts from (a) showing how the main and dimer peaks evolve as a function of $p_a$. (c)~Amplitude and (d) correlation length $\xi$ of the peaks as a function of $p_a$.}
\end{figure}

    \begin{table}[t]
    \centering
    \renewcommand{\arraystretch}{1.0} 
    \begin{tabularx}{\columnwidth}{@{}lcc@{}}
        \toprule
        Experiment & ~~~~~~~\emph{l} (r.l.u.)~~~~~~~ & Corresponding \\
        & & dimer fraction $p_a$\\
        \midrule
        \midrule
        Lauhlé \textit{et al.} \cite{Laulhe2015} & 0.20 & 0.66 \\
        Stahl \textit{et al.} \cite{Stahl2020} &  &  \\
        \midrule
        Ritschel \textit{et al.} \cite{Ritschel2015} & 0.18 & 0.84  \\
        Nicholson \textit{et al.} \cite{Nicolson2024} & &\\
        \bottomrule
    \end{tabularx}
    \caption{Out-of-plane CCDW Bragg peak positions from previous X-ray diffraction studies and the corresponding material fraction that is dimerized according to our HT model.
    }
    \label{tab:hopping}
    \end{table}

First, we calculate the XRD intensity $I(q_{l})$ at fixed $(q_{h}, q_{k})$ matching the in-plane reciprocal space positions of the CCDW diffraction peaks for different stacking probability distributions. Here we set $N = 100$, corresponding to a typical flake thickness of $\approx$ 60 nm. Figure~2(a) shows  $I(q_{l})$ for \emph{T$_{c}$}~stacking and a distribution varying from all monolayers ($p_{a}$ = 0) to a fully dimerized state ($p_{a}$ = 1). Line cuts at selected $p_{a}$ are shown in Fig.~2(b) and illustrate how the main and dimer peaks evolve as a function of the degree of dimerization. Complete dimerization with \emph{T$_{c}$} stacking results in a main peak position at \emph{l} = $\frac{1}{6}$ r.l.u. This is not consistent with XRD experiments and indicates that the CCDW state of \mbox{$1T$-TaS$_2$} is not fully dimerized. Comparing with the peak positions found in XRD experiments (see Tab.~1), we can estimate the fraction of dimers in the system: The five-layer periodicity from $l \approx \frac{1}{5}$ r.l.u originates from $p_{a} \approx \frac{2}{3}$, corresponding to an average distribution of two dimerized layers to one monolayer. We also conclude that the bulk cannot contain significant $T_b$ stacking as the calculated diffraction  would deviate further from the experiments (see Supplementary Material).

\begin{figure}[t]
	\includegraphics[width=\columnwidth]{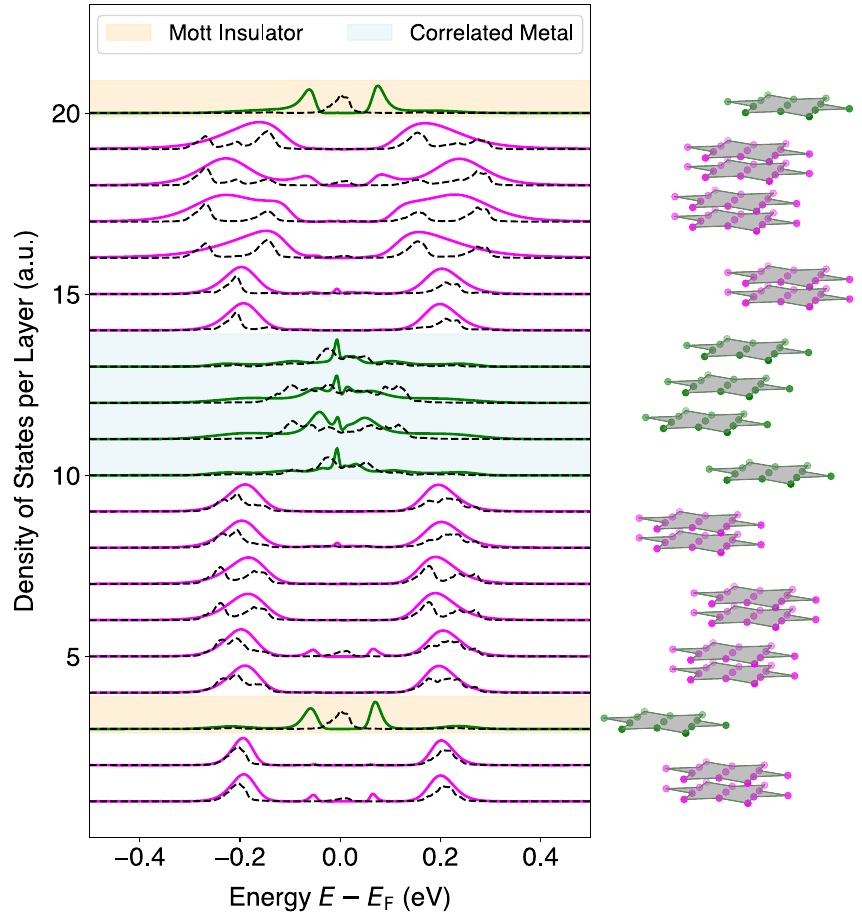}
	\caption{\label{Fig1}
	Black dashed lines correspond to the layer-resolved density of states (DOS), while green and pink lines depict the corresponding monolayer and dimer local spectral function~$A_n\left(\omega\right)$, respectively, calculated from DMFT for an exemplary 20-layer 1\emph{T}-TaS$_{2}$ system. Except for the dimers ($T_a$), the stacking between the layers is all $T_c$. The $\sim$ 0.4 eV wide gap, present already at the non-interacting level, stems from the strong hybridization of $T_a$ stacked layers. Mott insulating and correlated metallic layers are marked in yellow and green, respectively, while the unshaded layers are band insulating.
    }
\end{figure}

Next, we perform DMFT calculations on ten representative structures, each with a thickness of 20 layers, that follow a dimer-to-monolayer distribution of~2:1 and $T_{c}$~stacking. For each structure, we construct a low-energy Hamiltonian $H$ with a single orbital per layer. This orbital corresponds to the half-filled band with $d_{z^2}$~character that emerges in the low-temperature CCDW phase and is localized on the Ta site at the center of each polaron star \cite{LongYu2017, Lee2019}. The standard procedure to obtain $H$ involves performing DFT calculations for the desired lattice structure and projecting the resulting Kohn-Sham band structure onto a set of localized Wannier states with defined orbital character. This projection enables a tight-binding parametrization of the effective Hamiltonian. However, the computational cost increases rapidly with the number of sites in the unit cell, which would include 780 atoms for our setup. We therefore adopt a simpler, though well-motivated procedure: the effective single-band tight-binding Hamiltonian for the $n^{\rm th}$ isolated monolayer is calculated by DFT, whereas the real-space structure of the hoppings between layers is determined by the stacking arrangement of the adjacent layers. Since one effective orbital is associated with each polaron star, a single hopping parameter $t_v$, connecting the polarons on neighboring layers, is sufficient to account for the stacking arrangement. This procedure was also previously employed~\cite{Petocchi2022,Nicolson2024} where the $t_v$ magnitudes, associated with the $T_a$ and $T_c$ configurations, were optimized against STM measurements. While this minimal model may not describe all details of the multilayer system's electronic structure, it qualitatively captures how the kinetic energy is affected by the specific stacking arrangement. Importantly, our modeling reproduces the hybridization gap associated with the $T_a$ configuration already in the absence of correlations (see the dashed layer-resolved spectra in Fig.~3). In the  simulations, we use a Hubbard repulsion of $U=0.2$~eV~\cite{Petocchi2022,Nicolson2024} and account for local electronic correlations with a real-space extension of DMFT that provides an independent local self-energy for the $n^{\rm th}$ layer. All DMFT calculations are performed assuming $T\sim 30$~K and a half-filled system. Layer-resolved spectral functions $A_n\left(\omega\right)$ are obtained with the maximum entropy method. We use open boundary conditions in the stacking direction, so the external layers are representative of surface states.

Correlation effects are small in dimerized layers, as can be seen from the pink curves in Fig.~3, which resemble the non-interacting results in black. The situation is different for non-dimerized layers connected by \emph{T$_{c}$} stacking. In the case of monolayers surrounded exclusively by dimers, we observe a gap in the spectrum  only when using DMFT, \textit{i.e.} when electron correlations are considered, and the diverging self-energy at low Matsubara frequencies demonstrates the Mott insulating nature of the state (yellow shading). On the other hand, when stackings different from $T_a$ occur across a couple of layers, a strongly correlated metallic state emerges within the bulk (green shading). 

\begin{figure}[t]
	\includegraphics[width=0.98\columnwidth]{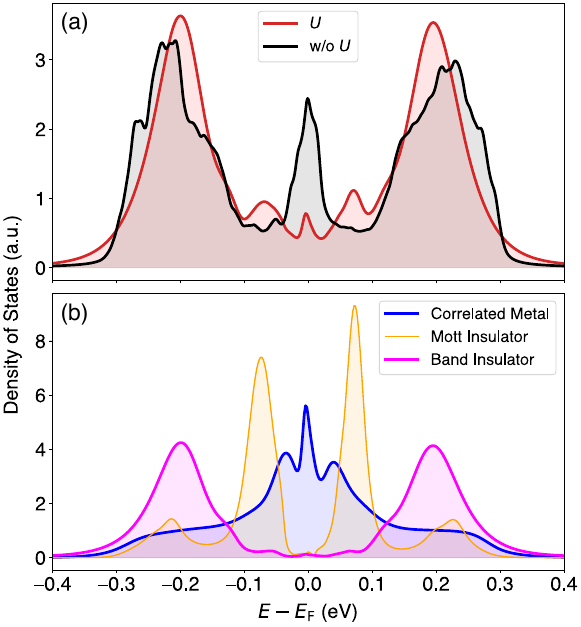}
	\caption{\label{Fig4}
	(a) Average spectral function $A(\omega)$ from ten 20-layer 1\emph{T}-TaS$_{2}$ systems with and without the local Hubbard \emph{U} interaction term. (b) $A_n(\omega)$ averaged over metallic (24 layers), Mott (14 layers) and band (162~layers) insulating states.}
\end{figure}

Taking all ten configurations into account, we compute the quantity most relevant for electrical transport in devices~\cite{Stojchevska2014,Burri2024}, namely the average spectral function $A(\omega)$, shown in Fig.~\ref{Fig4}(a). While the quasi-particle peak of the non-interacting DOS~(black) is suppressed when accounting for $U$ and a noticeable $\sim0.4$~eV gap is present (red), finite spectral weight remains at the Fermi energy $E_{\rm F}$. By analyzing the divergence of the self-energy, we categorize the layers by their electronic behaviors: correlated metal, as well as Mott and band insulators. The spectral functions for these states are shown in Fig. \ref{Fig4}(b) revealing that the non-zero spectral weight at $E_{\rm F}$ is due to the presence of the correlated metallic layers. Thus, we find that all dimerized layers are band insulating. Isolated monolayers tend to be Mott insulating while consecutive monolayers are metallic, although, the precise local configuration matters (see Supplementary Material).

In summary, we employ a recursive variation of the HT~method~\cite{Hendricks1942, Treacy1991} to exhaust the phase space of possible stacking configurations, including the degree of dimerization. Inputting these configurations into structure factor calculations to simulate the X-ray diffraction intensities, we conclude that the low-temperature, ``insulating" CCDW state of 1\emph{T}-TaS$_{2}$ is dominated by \emph{T$_{c}$} stacking and a mix of \emph{T$_a$}-stacked dimers and monolayers in a ratio of $\approx2:1$, where the dimers are band insulating, and---from our DMFT calculations---the monolayers behave as strongly correlated metals or Mott insulators depending upon their stacking neighborhood. Earlier studies suggested that the cooling rate to the CCDW state can affect the degree of dimerization \cite{Lee2019, delaTorre2025} and, hence, the material's electronic properties. Our results substantiate this possibility based on the variation of the out-of-plane CCDW diffraction peak position observed in different XRD experiments~\cite{Laulhe2015, Ritschel2015, Stahl2020, Nicolson2024}. Furthermore, the confirmation of buried Mott insulating monolayers identifies  hosts for possible QSL states  \cite{Law2017, Kratochvilova2017, Manas-Valero2021, Bozin2023} in this stacked two-dimensional material. The formalism we present here is therefore a powerful tool that connects XRD data to the real-space interlayer stacking structure, and which can be extended to other equilibrium and non-equilibrium phases of 1\emph{T}-TaS$_{2}$ (see End Matter), as well as a wider range of vdW materials \cite{Ren2022,  Ekahana2024, Venturini2024} with different stacking configurations. 
 
Our DMFT calculations  reveal the coexistence of correlated metallic and both band and Mott insulating layers in 1\emph{T}-TaS$_{2}$. This explains the conflicting results seen with surface-sensitive techniques, such as ARPES and STM that depend on the exfoliated surface layer. Through this reconciliation, we also demonstrate how interlayer hopping together with on-site Coulomb interactions controls the electronic properties of randomly stacked vdW materials. While fixed heterostructures with gates and Moiré or translated layers have provided very fertile ground for the study of many-body physics, thick vdW flakes of a single material represent a more accessible starting point for \emph{in-situ} tuning of the stacking, e.g. by thermal protocols or weak electrical and optical pulses. This approach has recently been explored to access nonequilibrium states not only for \mbox{1\emph{T}-TaS$_{2}$} but also other vdW systems, such as EuTe$_{4}$ \cite{Venturini2024}, or to introduce interlayer sliding for ferroelectric polarization control \cite{Wu2021, Bian2024, Yasuda2024,Stern2025}. The averaging implicit for thicker flakes should lead to greater reproducibility, and therefore greater application potential, for the random stacks characteristic of vdW materials.  The combination of the HT method with DMFT calculations introduced here represents a significant advance as it provides a framework to account for the structural and electronic behaviors associated with exactly such stacks.

\section{Acknowledgements}

We acknowledge fruitful discussions with D. Mihailovic. This research was funded by the Swiss National Science Foundation and the Slovenian Research And Innovation Agency as a part of the WEAVE framework Grant Number 213148. G.A. acknowledges funding from the European Research Council under the European Union’s Horizon 2020 Research and Innovation Programme, within Grant Agreement~810451 (HERO).

\bibliography{apssamp}

\providecommand{\noopsort}[1]{}\providecommand{\singleletter}[1]{#1}%
\begin{thebibliography}{59}%
\makeatletter
\providecommand \@ifxundefined [1]{%
 \@ifx{#1\undefined}
}%
\providecommand \@ifnum [1]{%
 \ifnum #1\expandafter \@firstoftwo
 \else \expandafter \@secondoftwo
 \fi
}%
\providecommand \@ifx [1]{%
 \ifx #1\expandafter \@firstoftwo
 \else \expandafter \@secondoftwo
 \fi
}%
\providecommand \natexlab [1]{#1}%
\providecommand \enquote  [1]{``#1''}%
\providecommand \bibnamefont  [1]{#1}%
\providecommand \bibfnamefont [1]{#1}%
\providecommand \citenamefont [1]{#1}%
\providecommand \href@noop [0]{\@secondoftwo}%
\providecommand \href [0]{\begingroup \@sanitize@url \@href}%
\providecommand \@href[1]{\@@startlink{#1}\@@href}%
\providecommand \@@href[1]{\endgroup#1\@@endlink}%
\providecommand \@sanitize@url [0]{\catcode `\\12\catcode `\$12\catcode
  `\&12\catcode `\#12\catcode `\^12\catcode `\_12\catcode `\%12\relax}%
\providecommand \@@startlink[1]{}%
\providecommand \@@endlink[0]{}%
\providecommand \url  [0]{\begingroup\@sanitize@url \@url }%
\providecommand \@url [1]{\endgroup\@href {#1}{\urlprefix }}%
\providecommand \urlprefix  [0]{URL }%
\providecommand \Eprint [0]{\href }%
\providecommand \doibase [0]{https://doi.org/}%
\providecommand \selectlanguage [0]{\@gobble}%
\providecommand \bibinfo  [0]{\@secondoftwo}%
\providecommand \bibfield  [0]{\@secondoftwo}%
\providecommand \translation [1]{[#1]}%
\providecommand \BibitemOpen [0]{}%
\providecommand \bibitemStop [0]{}%
\providecommand \bibitemNoStop [0]{.\EOS\space}%
\providecommand \EOS [0]{\spacefactor3000\relax}%
\providecommand \BibitemShut  [1]{\csname bibitem#1\endcsname}%
\let\auto@bib@innerbib\@empty
\bibitem [{\citenamefont {Qian}\ \emph {et~al.}(2014)\citenamefont {Qian},
  \citenamefont {Liu}, \citenamefont {Fu},\ and\ \citenamefont
  {Li}}]{Qian2014}%
  \BibitemOpen
  \bibfield  {author} {\bibinfo {author} {\bibfnamefont {X.}~\bibnamefont
  {Qian}}, \bibinfo {author} {\bibfnamefont {J.}~\bibnamefont {Liu}}, \bibinfo
  {author} {\bibfnamefont {L.}~\bibnamefont {Fu}},\ and\ \bibinfo {author}
  {\bibfnamefont {J.}~\bibnamefont {Li}},\ }\bibfield  {title} {\bibinfo
  {title} {Quantum spin {Hall} effect in two-dimensional transition metal
  dichalcogenides},\ }\href@noop {} {\bibfield  {journal} {\bibinfo  {journal}
  {Science}\ }\textbf {\bibinfo {volume} {346}},\ \bibinfo {pages} {1344}
  (\bibinfo {year} {2014})}\BibitemShut {NoStop}%
\bibitem [{\citenamefont {Fei}\ \emph {et~al.}(2017)\citenamefont {Fei} \emph
  {et~al.}}]{Fei2017}%
  \BibitemOpen
  \bibfield  {author} {\bibinfo {author} {\bibfnamefont {Z.}~\bibnamefont
  {Fei}} \emph {et~al.},\ }\bibfield  {title} {\bibinfo {title} {Edge
  conduction in monolayer {${\mathrm{WTe}}_{2}$}},\ }\href@noop {} {\bibfield
  {journal} {\bibinfo  {journal} {Nat.~Phys.}\ }\textbf {\bibinfo {volume}
  {13}},\ \bibinfo {pages} {677} (\bibinfo {year} {2017})}\BibitemShut
  {NoStop}%
\bibitem [{\citenamefont {Geim}\ and\ \citenamefont
  {Grigorieva}(2013)}]{Geim2013}%
  \BibitemOpen
  \bibfield  {author} {\bibinfo {author} {\bibfnamefont {A.~K.}\ \bibnamefont
  {Geim}}\ and\ \bibinfo {author} {\bibfnamefont {I.~V.}\ \bibnamefont
  {Grigorieva}},\ }\bibfield  {title} {\bibinfo {title} {{Van der Waals}
  heterostructures},\ }\href@noop {} {\bibfield  {journal} {\bibinfo  {journal}
  {Nature}\ }\textbf {\bibinfo {volume} {499}},\ \bibinfo {pages} {419}
  (\bibinfo {year} {2013})}\BibitemShut {NoStop}%
\bibitem [{\citenamefont {Ould~NE}\ \emph {et~al.}(2017)\citenamefont
  {Ould~NE}, \citenamefont {Boujnah}, \citenamefont {Benyoussef},\ and\
  \citenamefont {El~Kenz}}]{Ould2017}%
  \BibitemOpen
  \bibfield  {author} {\bibinfo {author} {\bibfnamefont {M.~L.}\ \bibnamefont
  {Ould~NE}}, \bibinfo {author} {\bibfnamefont {M.}~\bibnamefont {Boujnah}},
  \bibinfo {author} {\bibfnamefont {A.}~\bibnamefont {Benyoussef}},\ and\
  \bibinfo {author} {\bibfnamefont {A.}~\bibnamefont {El~Kenz}},\ }\bibfield
  {title} {\bibinfo {title} {Electronic and electrical conductivity of {AB} and
  {AA}-stacked bilayer graphene with tunable layer separation},\ }\href@noop {}
  {\bibfield  {journal} {\bibinfo  {journal} {J. Supercond. Nov. Magn.}\
  }\textbf {\bibinfo {volume} {30}},\ \bibinfo {pages} {1263} (\bibinfo {year}
  {2017})}\BibitemShut {NoStop}%
\bibitem [{\citenamefont {Liang}\ \emph {et~al.}(2020)\citenamefont {Liang}
  \emph {et~al.}}]{Liang2020}%
  \BibitemOpen
  \bibfield  {author} {\bibinfo {author} {\bibfnamefont {X.}~\bibnamefont
  {Liang}} \emph {et~al.},\ }\bibfield  {title} {\bibinfo {title} {Effect of
  bilayer stacking on the atomic and electronic structure of twisted double
  bilayer graphene},\ }\href@noop {} {\bibfield  {journal} {\bibinfo  {journal}
  {Phys. Rev. B}\ }\textbf {\bibinfo {volume} {102}},\ \bibinfo {pages}
  {155146} (\bibinfo {year} {2020})}\BibitemShut {NoStop}%
\bibitem [{\citenamefont {Bistritzer}\ and\ \citenamefont
  {MacDonald}(2011)}]{Bistritzer2011}%
  \BibitemOpen
  \bibfield  {author} {\bibinfo {author} {\bibfnamefont {R.}~\bibnamefont
  {Bistritzer}}\ and\ \bibinfo {author} {\bibfnamefont {A.~H.}\ \bibnamefont
  {MacDonald}},\ }\bibfield  {title} {\bibinfo {title} {Moiré bands in twisted
  double-layer graphene},\ }\href@noop {} {\bibfield  {journal} {\bibinfo
  {journal} {Proc. Natl. Acad. Sci. USA}\ }\textbf {\bibinfo {volume} {108}},\
  \bibinfo {pages} {12233} (\bibinfo {year} {2011})}\BibitemShut {NoStop}%
\bibitem [{\citenamefont {Cao}\ \emph {et~al.}(2018)\citenamefont {Cao} \emph
  {et~al.}}]{Cao2018}%
  \BibitemOpen
  \bibfield  {author} {\bibinfo {author} {\bibfnamefont {Y.}~\bibnamefont
  {Cao}} \emph {et~al.},\ }\bibfield  {title} {\bibinfo {title} {Unconventional
  superconductivity in magic-angle graphene superlattices},\ }\href@noop {}
  {\bibfield  {journal} {\bibinfo  {journal} {Nature}\ }\textbf {\bibinfo
  {volume} {556}},\ \bibinfo {pages} {43} (\bibinfo {year} {2018})}\BibitemShut
  {NoStop}%
\bibitem [{\citenamefont {Liu}\ \emph {et~al.}(2016)\citenamefont {Liu} \emph
  {et~al.}}]{Liu2016}%
  \BibitemOpen
  \bibfield  {author} {\bibinfo {author} {\bibfnamefont {Y.}~\bibnamefont
  {Liu}} \emph {et~al.},\ }\bibfield  {title} {\bibinfo {title} {Van der
  {W}aals heterostructures and devices},\ }\href@noop {} {\bibfield  {journal}
  {\bibinfo  {journal} {Nat. Rev. Mater.}\ }\textbf {\bibinfo {volume} {1}},\
  \bibinfo {pages} {16042} (\bibinfo {year} {2016})}\BibitemShut {NoStop}%
\bibitem [{\citenamefont {Wang}\ \emph
  {et~al.}(2023{\natexlab{a}})\citenamefont {Wang}, \citenamefont {Zhao},
  \citenamefont {Ming},\ and\ \citenamefont {Si}}]{Wang2023_1T-TaSe2}%
  \BibitemOpen
  \bibfield  {author} {\bibinfo {author} {\bibfnamefont {W.}~\bibnamefont
  {Wang}}, \bibinfo {author} {\bibfnamefont {B.}~\bibnamefont {Zhao}}, \bibinfo
  {author} {\bibfnamefont {X.}~\bibnamefont {Ming}},\ and\ \bibinfo {author}
  {\bibfnamefont {C.}~\bibnamefont {Si}},\ }\bibfield  {title} {\bibinfo
  {title} {Multiple quantum states induced in {1\emph{T}-TaSe$_{2}$} by
  controlling the stacking order of charge density waves},\ }\href@noop {}
  {\bibfield  {journal} {\bibinfo  {journal} {Adv. Funct. Mater.}\ }\textbf
  {\bibinfo {volume} {33}},\ \bibinfo {pages} {2214583} (\bibinfo {year}
  {2023}{\natexlab{a}})}\BibitemShut {NoStop}%
\bibitem [{\citenamefont {Stojchevska}\ \emph {et~al.}(2014)\citenamefont
  {Stojchevska} \emph {et~al.}}]{Stojchevska2014}%
  \BibitemOpen
  \bibfield  {author} {\bibinfo {author} {\bibfnamefont {L.}~\bibnamefont
  {Stojchevska}} \emph {et~al.},\ }\bibfield  {title} {\bibinfo {title}
  {Ultrafast switching to a stable hidden quantum state in an electronic
  crystal},\ }\href@noop {} {\bibfield  {journal} {\bibinfo  {journal}
  {Science}\ }\textbf {\bibinfo {volume} {344}},\ \bibinfo {pages} {177}
  (\bibinfo {year} {2014})}\BibitemShut {NoStop}%
\bibitem [{\citenamefont {Wang}\ \emph {et~al.}(2012)\citenamefont {Wang} \emph
  {et~al.}}]{Wang2012}%
  \BibitemOpen
  \bibfield  {author} {\bibinfo {author} {\bibfnamefont {Q.~H.}\ \bibnamefont
  {Wang}} \emph {et~al.},\ }\bibfield  {title} {\bibinfo {title} {Electronics
  and optoelectronics of two-dimensional transition metal dichalcogenides},\
  }\href@noop {} {\bibfield  {journal} {\bibinfo  {journal} {Nat.
  Nanotechnol.}\ }\textbf {\bibinfo {volume} {7}},\ \bibinfo {pages} {699}
  (\bibinfo {year} {2012})}\BibitemShut {NoStop}%
\bibitem [{\citenamefont {Jariwala}\ \emph {et~al.}(2014)\citenamefont
  {Jariwala} \emph {et~al.}}]{Jariwala2014}%
  \BibitemOpen
  \bibfield  {author} {\bibinfo {author} {\bibfnamefont {D.}~\bibnamefont
  {Jariwala}} \emph {et~al.},\ }\bibfield  {title} {\bibinfo {title} {Emerging
  device applications for semiconducting two-dimensional transition metal
  dichalcogenides},\ }\href@noop {} {\bibfield  {journal} {\bibinfo  {journal}
  {ACS Nano}\ }\textbf {\bibinfo {volume} {8}},\ \bibinfo {pages} {1102}
  (\bibinfo {year} {2014})}\BibitemShut {NoStop}%
\bibitem [{\citenamefont {Stahl}\ \emph {et~al.}(2020)\citenamefont {Stahl}
  \emph {et~al.}}]{Stahl2020}%
  \BibitemOpen
  \bibfield  {author} {\bibinfo {author} {\bibfnamefont {Q.}~\bibnamefont
  {Stahl}} \emph {et~al.},\ }\bibfield  {title} {\bibinfo {title} {Collapse of
  layer dimerization in the photo-induced hidden state of
  {1\emph{T}-TaS$_{2}$}},\ }\href@noop {} {\bibfield  {journal} {\bibinfo
  {journal} {Nat. Commun.}\ }\textbf {\bibinfo {volume} {11}},\ \bibinfo
  {pages} {1247} (\bibinfo {year} {2020})}\BibitemShut {NoStop}%
\bibitem [{\citenamefont {Burri}\ \emph {et~al.}()\citenamefont {Burri} \emph
  {et~al.}}]{Burri2024}%
  \BibitemOpen
  \bibfield  {author} {\bibinfo {author} {\bibfnamefont {C.}~\bibnamefont
  {Burri}} \emph {et~al.},\ }\href@noop {} {\bibinfo {title} {Imaging of
  electrically controlled van der {W}aals layer stacking in
  {1\emph{T}-TaS$_{2}$}}},\ \Eprint {https://arxiv.org/abs/2411.04830}
  {arXiv:2411.04830} \BibitemShut {NoStop}%
\bibitem [{\citenamefont {Tsen}\ \emph {et~al.}(2015)\citenamefont {Tsen} \emph
  {et~al.}}]{Tsen2015}%
  \BibitemOpen
  \bibfield  {author} {\bibinfo {author} {\bibfnamefont {A.~W.}\ \bibnamefont
  {Tsen}} \emph {et~al.},\ }\bibfield  {title} {\bibinfo {title} {Structure and
  control of charge density waves in two-dimensional
  1\emph{T}-{${\mathrm{TaS}}_{2}$}},\ }\href@noop {} {\bibfield  {journal}
  {\bibinfo  {journal} {Proc. Natl. Acad. Sci. USA}\ }\textbf {\bibinfo
  {volume} {112}},\ \bibinfo {pages} {15054} (\bibinfo {year}
  {2015})}\BibitemShut {NoStop}%
\bibitem [{\citenamefont {Wang}\ \emph
  {et~al.}(2023{\natexlab{b}})\citenamefont {Wang} \emph {et~al.}}]{Wang2023}%
  \BibitemOpen
  \bibfield  {author} {\bibinfo {author} {\bibfnamefont {G.}~\bibnamefont
  {Wang}} \emph {et~al.},\ }\bibfield  {title} {\bibinfo {title} {Atomic
  visualization of the {3D} charge density wave stacking in
  {1\emph{T}-TaS$_{2}$} by cryogenic transmission electron microscopy},\
  }\href@noop {} {\bibfield  {journal} {\bibinfo  {journal} {Nano Lett.}\
  }\textbf {\bibinfo {volume} {23}},\ \bibinfo {pages} {4318} (\bibinfo {year}
  {2023}{\natexlab{b}})}\BibitemShut {NoStop}%
\bibitem [{\citenamefont {Ma}\ \emph {et~al.}(2016)\citenamefont {Ma} \emph
  {et~al.}}]{Ma2016}%
  \BibitemOpen
  \bibfield  {author} {\bibinfo {author} {\bibfnamefont {L.}~\bibnamefont {Ma}}
  \emph {et~al.},\ }\bibfield  {title} {\bibinfo {title} {A metallic mosaic
  phase and the origin of {M}ott-insulating state in {1\emph{T}-TaS$_{2}$}},\
  }\href@noop {} {\bibfield  {journal} {\bibinfo  {journal} {Nat. Commun.}\
  }\textbf {\bibinfo {volume} {7}},\ \bibinfo {pages} {10956} (\bibinfo {year}
  {2016})}\BibitemShut {NoStop}%
\bibitem [{\citenamefont {Gerasimenko}\ \emph {et~al.}(2019)\citenamefont
  {Gerasimenko}, \citenamefont {Karpov}, \citenamefont {Vaskivskyi},
  \citenamefont {Brazovskii},\ and\ \citenamefont
  {Mihailovic}}]{Gerasimenko2019}%
  \BibitemOpen
  \bibfield  {author} {\bibinfo {author} {\bibfnamefont {Y.~A.}\ \bibnamefont
  {Gerasimenko}}, \bibinfo {author} {\bibfnamefont {P.}~\bibnamefont {Karpov}},
  \bibinfo {author} {\bibfnamefont {I.}~\bibnamefont {Vaskivskyi}}, \bibinfo
  {author} {\bibfnamefont {S.}~\bibnamefont {Brazovskii}},\ and\ \bibinfo
  {author} {\bibfnamefont {D.}~\bibnamefont {Mihailovic}},\ }\bibfield  {title}
  {\bibinfo {title} {Intertwined chiral charge orders and topological
  stabilization of the light-induced state of a prototypical transition metal
  dichalcogenide},\ }\href@noop {} {\bibfield  {journal} {\bibinfo  {journal}
  {npj Quantum Mater.}\ }\textbf {\bibinfo {volume} {4}},\ \bibinfo {pages}
  {32} (\bibinfo {year} {2019})}\BibitemShut {NoStop}%
\bibitem [{\citenamefont {Ravnik}\ \emph {et~al.}(2021)\citenamefont {Ravnik}
  \emph {et~al.}}]{Ravnik2021}%
  \BibitemOpen
  \bibfield  {author} {\bibinfo {author} {\bibfnamefont {J.}~\bibnamefont
  {Ravnik}} \emph {et~al.},\ }\bibfield  {title} {\bibinfo {title} {A
  time-domain phase diagram of metastable states in a charge ordered quantum
  material},\ }\href@noop {} {\bibfield  {journal} {\bibinfo  {journal} {Nat.
  Commun.}\ }\textbf {\bibinfo {volume} {12}},\ \bibinfo {pages} {2323}
  (\bibinfo {year} {2021})}\BibitemShut {NoStop}%
\bibitem [{\citenamefont {Tanda}\ \emph {et~al.}(1984)\citenamefont {Tanda},
  \citenamefont {Sambongi}, \citenamefont {Tani},\ and\ \citenamefont
  {Tanaka}}]{Tanda1984}%
  \BibitemOpen
  \bibfield  {author} {\bibinfo {author} {\bibfnamefont {S.}~\bibnamefont
  {Tanda}}, \bibinfo {author} {\bibfnamefont {T.}~\bibnamefont {Sambongi}},
  \bibinfo {author} {\bibfnamefont {T.}~\bibnamefont {Tani}},\ and\ \bibinfo
  {author} {\bibfnamefont {S.}~\bibnamefont {Tanaka}},\ }\bibfield  {title}
  {\bibinfo {title} {\mbox{X-Ray} study of charge density wave structure in
  {1\emph{T}-TaS$_{2}$}},\ }\href@noop {} {\bibfield  {journal} {\bibinfo
  {journal} {J.~Phys. Soc. Jpn.}\ }\textbf {\bibinfo {volume} {53}},\ \bibinfo
  {pages} {476} (\bibinfo {year} {1984})}\BibitemShut {NoStop}%
\bibitem [{\citenamefont {Nakanishi}\ and\ \citenamefont
  {Shiba}(1984)}]{Nakanishi1984}%
  \BibitemOpen
  \bibfield  {author} {\bibinfo {author} {\bibfnamefont {K.}~\bibnamefont
  {Nakanishi}}\ and\ \bibinfo {author} {\bibfnamefont {H.}~\bibnamefont
  {Shiba}},\ }\bibfield  {title} {\bibinfo {title} {Theory of three-dimensional
  orderings of charge-density waves in {1\emph{T}-TaX$_{2}$} {(X: S, Se)}},\
  }\href@noop {} {\bibfield  {journal} {\bibinfo  {journal} {J. Phys. Soc.
  Jpn.}\ }\textbf {\bibinfo {volume} {53}},\ \bibinfo {pages} {1103} (\bibinfo
  {year} {1984})}\BibitemShut {NoStop}%
\bibitem [{\citenamefont {Butler}\ \emph {et~al.}(2020)\citenamefont {Butler},
  \citenamefont {Yoshida}, \citenamefont {Hanaguri},\ and\ \citenamefont
  {Iwasa}}]{Butler2020}%
  \BibitemOpen
  \bibfield  {author} {\bibinfo {author} {\bibfnamefont {C.~J.}\ \bibnamefont
  {Butler}}, \bibinfo {author} {\bibfnamefont {M.}~\bibnamefont {Yoshida}},
  \bibinfo {author} {\bibfnamefont {T.}~\bibnamefont {Hanaguri}},\ and\
  \bibinfo {author} {\bibfnamefont {Y.}~\bibnamefont {Iwasa}},\ }\bibfield
  {title} {\bibinfo {title} {Mottness versus unit-cell doubling as the driver
  of the insulating state in {1\emph{T}-TaS$_{2}$}},\ }\href@noop {} {\bibfield
   {journal} {\bibinfo  {journal} {Nat. Commun.}\ }\textbf {\bibinfo {volume}
  {11}},\ \bibinfo {pages} {2477} (\bibinfo {year} {2020})}\BibitemShut
  {NoStop}%
\bibitem [{\citenamefont {Perfetti}\ \emph {et~al.}(2006)\citenamefont
  {Perfetti} \emph {et~al.}}]{Perfetti2006}%
  \BibitemOpen
  \bibfield  {author} {\bibinfo {author} {\bibfnamefont {L.}~\bibnamefont
  {Perfetti}} \emph {et~al.},\ }\bibfield  {title} {\bibinfo {title} {Time
  evolution of the electronic structure of {1\emph{T}-TaS$_{2}$} through the
  insulator-metal transition},\ }\href@noop {} {\bibfield  {journal} {\bibinfo
  {journal} {Phys. Rev. Lett.}\ }\textbf {\bibinfo {volume} {97}},\ \bibinfo
  {pages} {067402} (\bibinfo {year} {2006})}\BibitemShut {NoStop}%
\bibitem [{\citenamefont {Hellmann}\ \emph {et~al.}(2010)\citenamefont
  {Hellmann} \emph {et~al.}}]{Hellmann2010}%
  \BibitemOpen
  \bibfield  {author} {\bibinfo {author} {\bibfnamefont {S.}~\bibnamefont
  {Hellmann}} \emph {et~al.},\ }\bibfield  {title} {\bibinfo {title} {Ultrafast
  melting of a charge-density wave in the {M}ott insulator
  {1\emph{T}-TaS$_{2}$}},\ }\href@noop {} {\bibfield  {journal} {\bibinfo
  {journal} {Phys. Rev. Lett.}\ }\textbf {\bibinfo {volume} {105}},\ \bibinfo
  {pages} {187401} (\bibinfo {year} {2010})}\BibitemShut {NoStop}%
\bibitem [{\citenamefont {Hellmann}\ \emph {et~al.}(2012)\citenamefont
  {Hellmann} \emph {et~al.}}]{Hellmann2012}%
  \BibitemOpen
  \bibfield  {author} {\bibinfo {author} {\bibfnamefont {S.}~\bibnamefont
  {Hellmann}} \emph {et~al.},\ }\bibfield  {title} {\bibinfo {title}
  {Time-domain classification of charge-density-wave insulators},\ }\href@noop
  {} {\bibfield  {journal} {\bibinfo  {journal} {Nat. Commun.}\ }\textbf
  {\bibinfo {volume} {3}},\ \bibinfo {pages} {1069} (\bibinfo {year}
  {2012})}\BibitemShut {NoStop}%
\bibitem [{\citenamefont {Ritschel}\ \emph {et~al.}()\citenamefont {Ritschel}
  \emph {et~al.}}]{Ritschel2015}%
  \BibitemOpen
  \bibfield  {author} {\bibinfo {author} {\bibfnamefont {T.}~\bibnamefont
  {Ritschel}} \emph {et~al.},\ }\href@noop {} {\ }\BibitemShut {NoStop}%
\bibitem [{\citenamefont {Wang}\ \emph {et~al.}(2020)\citenamefont {Wang} \emph
  {et~al.}}]{Wang2020}%
  \BibitemOpen
  \bibfield  {author} {\bibinfo {author} {\bibfnamefont {Y.~D.}\ \bibnamefont
  {Wang}} \emph {et~al.},\ }\bibfield  {title} {\bibinfo {title} {Band
  insulator to {M}ott insulator transition in {1\emph{T}-TaS$_{2}$}},\
  }\href@noop {} {\bibfield  {journal} {\bibinfo  {journal} {Nat. Commun.}\
  }\textbf {\bibinfo {volume} {11}},\ \bibinfo {pages} {4215} (\bibinfo {year}
  {2020})}\BibitemShut {NoStop}%
\bibitem [{\citenamefont {Yang}\ \emph {et~al.}(2024)\citenamefont {Yang} \emph
  {et~al.}}]{Yang2024}%
  \BibitemOpen
  \bibfield  {author} {\bibinfo {author} {\bibfnamefont {H.}~\bibnamefont
  {Yang}} \emph {et~al.},\ }\bibfield  {title} {\bibinfo {title} {Origin of
  distinct insulating domains in the layered charge density wave material
  {1\emph{T}-TaS$_{2}$}},\ }\href@noop {} {\bibfield  {journal} {\bibinfo
  {journal} {Adv.~Sci.}\ }\textbf {\bibinfo {volume} {11}},\ \bibinfo {pages}
  {2401348} (\bibinfo {year} {2024})}\BibitemShut {NoStop}%
\bibitem [{\citenamefont {Wang}\ \emph {et~al.}(2024)\citenamefont {Wang} \emph
  {et~al.}}]{Wang2024}%
  \BibitemOpen
  \bibfield  {author} {\bibinfo {author} {\bibfnamefont {Y.}~\bibnamefont
  {Wang}} \emph {et~al.},\ }\bibfield  {title} {\bibinfo {title} {Dualistic
  insulator states in {1\emph{T}-TaS$_{2}$} crystals},\ }\href@noop {}
  {\bibfield  {journal} {\bibinfo  {journal} {Nat. Commun.}\ }\textbf {\bibinfo
  {volume} {15}},\ \bibinfo {pages} {3425} (\bibinfo {year}
  {2024})}\BibitemShut {NoStop}%
\bibitem [{\citenamefont {Sugai}\ \emph {et~al.}(1981)\citenamefont {Sugai},
  \citenamefont {Murase}, \citenamefont {Uchida},\ and\ \citenamefont
  {Tanaka}}]{Sugai1981}%
  \BibitemOpen
  \bibfield  {author} {\bibinfo {author} {\bibfnamefont {S.}~\bibnamefont
  {Sugai}}, \bibinfo {author} {\bibfnamefont {K.}~\bibnamefont {Murase}},
  \bibinfo {author} {\bibfnamefont {S.}~\bibnamefont {Uchida}},\ and\ \bibinfo
  {author} {\bibfnamefont {S.}~\bibnamefont {Tanaka}},\ }\bibfield  {title}
  {\bibinfo {title} {Comparison of the soft modes in tantalum
  dichalcogenides},\ }\href@noop {} {\bibfield  {journal} {\bibinfo  {journal}
  {Physica B+C}\ }\textbf {\bibinfo {volume} {105}},\ \bibinfo {pages} {405}
  (\bibinfo {year} {1981})}\BibitemShut {NoStop}%
\bibitem [{\citenamefont {Toda}\ \emph {et~al.}(2004)\citenamefont {Toda},
  \citenamefont {Tateishi},\ and\ \citenamefont {Tanda}}]{Toda2004}%
  \BibitemOpen
  \bibfield  {author} {\bibinfo {author} {\bibfnamefont {Y.}~\bibnamefont
  {Toda}}, \bibinfo {author} {\bibfnamefont {K.}~\bibnamefont {Tateishi}},\
  and\ \bibinfo {author} {\bibfnamefont {S.}~\bibnamefont {Tanda}},\ }\bibfield
   {title} {\bibinfo {title} {Anomalous coherent phonon oscillations in the
  commensurate phase of the quasi-two-dimensional {1\emph{T}-TaS$_{2}$}
  compound},\ }\href@noop {} {\bibfield  {journal} {\bibinfo  {journal} {Phys.
  Rev.~B}\ }\textbf {\bibinfo {volume} {70}},\ \bibinfo {pages} {033106}
  (\bibinfo {year} {2004})}\BibitemShut {NoStop}%
\bibitem [{\citenamefont {Ravnik}\ \emph {et~al.}(2018)\citenamefont {Ravnik},
  \citenamefont {Vaskivskyi}, \citenamefont {Mertelj},\ and\ \citenamefont
  {Mihailovic}}]{Ravnik2017}%
  \BibitemOpen
  \bibfield  {author} {\bibinfo {author} {\bibfnamefont {J.}~\bibnamefont
  {Ravnik}}, \bibinfo {author} {\bibfnamefont {I.}~\bibnamefont {Vaskivskyi}},
  \bibinfo {author} {\bibfnamefont {T.}~\bibnamefont {Mertelj}},\ and\ \bibinfo
  {author} {\bibfnamefont {D.}~\bibnamefont {Mihailovic}},\ }\bibfield  {title}
  {\bibinfo {title} {Real-time observation of the coherent transition to a
  metastable emergent state in {1\emph{T}-TaS$_{2}$}},\ }\href@noop {}
  {\bibfield  {journal} {\bibinfo  {journal} {Phys. Rev. B}\ }\textbf {\bibinfo
  {volume} {97}},\ \bibinfo {pages} {075304} (\bibinfo {year}
  {2018})}\BibitemShut {NoStop}%
\bibitem [{\citenamefont {Tanimura}(2018)}]{Tanimura2018}%
  \BibitemOpen
  \bibfield  {author} {\bibinfo {author} {\bibfnamefont {K.}~\bibnamefont
  {Tanimura}},\ }\bibfield  {title} {\bibinfo {title} {Photoinduced
  discommensuration of the commensurate charge-density wave phase in
  {1\emph{T}-TaS$_{2}$}},\ }\href@noop {} {\bibfield  {journal} {\bibinfo
  {journal} {Phys. Rev. B}\ }\textbf {\bibinfo {volume} {97}},\ \bibinfo
  {pages} {245115} (\bibinfo {year} {2018})}\BibitemShut {NoStop}%
\bibitem [{\citenamefont {Albertini}\ \emph {et~al.}(2016)\citenamefont
  {Albertini} \emph {et~al.}}]{Albertini2016}%
  \BibitemOpen
  \bibfield  {author} {\bibinfo {author} {\bibfnamefont {O.~R.}\ \bibnamefont
  {Albertini}} \emph {et~al.},\ }\bibfield  {title} {\bibinfo {title}
  {Zone-center phonons of bulk, few-layer, and monolayer {1\emph{T}-TaS$_{2}$}:
  Detection of commensurate charge density wave phase through {R}aman
  scattering},\ }\href@noop {} {\bibfield  {journal} {\bibinfo  {journal}
  {Phys. Rev. B}\ }\textbf {\bibinfo {volume} {93}},\ \bibinfo {pages} {214109}
  (\bibinfo {year} {2016})}\BibitemShut {NoStop}%
\bibitem [{\citenamefont {Klanj{\v s}ek}\ \emph {et~al.}(2017)\citenamefont
  {Klanj{\v s}ek} \emph {et~al.}}]{Klanjvsek2017}%
  \BibitemOpen
  \bibfield  {author} {\bibinfo {author} {\bibfnamefont {M.}~\bibnamefont
  {Klanj{\v s}ek}} \emph {et~al.},\ }\bibfield  {title} {\bibinfo {title} {A
  high-temperature quantum spin liquid with polaron spins},\ }\href@noop {}
  {\bibfield  {journal} {\bibinfo  {journal} {Nat. Phys.}\ }\textbf {\bibinfo
  {volume} {13}},\ \bibinfo {pages} {1130} (\bibinfo {year}
  {2017})}\BibitemShut {NoStop}%
\bibitem [{\citenamefont {Law}\ and\ \citenamefont {Lee}(2017)}]{Law2017}%
  \BibitemOpen
  \bibfield  {author} {\bibinfo {author} {\bibfnamefont {K.~T.}\ \bibnamefont
  {Law}}\ and\ \bibinfo {author} {\bibfnamefont {P.~A.}\ \bibnamefont {Lee}},\
  }\bibfield  {title} {\bibinfo {title} {{1\emph{T}-TaS$_{2}$} as a quantum
  spin liquid},\ }\href@noop {} {\bibfield  {journal} {\bibinfo  {journal}
  {Proc. Natl. Acad. Sci. USA}\ }\textbf {\bibinfo {volume} {114}},\ \bibinfo
  {pages} {6996} (\bibinfo {year} {2017})}\BibitemShut {NoStop}%
\bibitem [{\citenamefont {Kratochvilova}\ \emph {et~al.}(2017)\citenamefont
  {Kratochvilova} \emph {et~al.}}]{Kratochvilova2017}%
  \BibitemOpen
  \bibfield  {author} {\bibinfo {author} {\bibfnamefont {M.}~\bibnamefont
  {Kratochvilova}} \emph {et~al.},\ }\bibfield  {title} {\bibinfo {title} {The
  low-temperature highly correlated quantum phase in the charge-density-wave
  \mbox{{1\emph{T}-TaS$_{2}$}} compound},\ }\href@noop {} {\bibfield  {journal}
  {\bibinfo  {journal} {npj Quantum Mater.}\ }\textbf {\bibinfo {volume} {2}},\
  \bibinfo {pages} {42} (\bibinfo {year} {2017})}\BibitemShut {NoStop}%
\bibitem [{\citenamefont {Ma{\~n}as-Valero}\ \emph {et~al.}(2021)\citenamefont
  {Ma{\~n}as-Valero}, \citenamefont {Huddart}, \citenamefont {Lancaster},
  \citenamefont {Coronado},\ and\ \citenamefont {Pratt}}]{Manas-Valero2021}%
  \BibitemOpen
  \bibfield  {author} {\bibinfo {author} {\bibfnamefont {S.}~\bibnamefont
  {Ma{\~n}as-Valero}}, \bibinfo {author} {\bibfnamefont {B.~M.}\ \bibnamefont
  {Huddart}}, \bibinfo {author} {\bibfnamefont {T.}~\bibnamefont {Lancaster}},
  \bibinfo {author} {\bibfnamefont {E.}~\bibnamefont {Coronado}},\ and\
  \bibinfo {author} {\bibfnamefont {F.~L.}\ \bibnamefont {Pratt}},\ }\bibfield
  {title} {\bibinfo {title} {Quantum phases and spin liquid properties of
  {1\emph{T}-TaS$_{2}$}},\ }\href@noop {} {\bibfield  {journal} {\bibinfo
  {journal} {npj Quantum Mater.}\ }\textbf {\bibinfo {volume} {6}},\ \bibinfo
  {pages} {69} (\bibinfo {year} {2021})}\BibitemShut {NoStop}%
\bibitem [{\citenamefont {Ritschel}\ \emph {et~al.}(2018)\citenamefont
  {Ritschel}, \citenamefont {Berger},\ and\ \citenamefont
  {Geck}}]{Ritschel2018}%
  \BibitemOpen
  \bibfield  {author} {\bibinfo {author} {\bibfnamefont {T.}~\bibnamefont
  {Ritschel}}, \bibinfo {author} {\bibfnamefont {H.}~\bibnamefont {Berger}},\
  and\ \bibinfo {author} {\bibfnamefont {J.}~\bibnamefont {Geck}},\ }\bibfield
  {title} {\bibinfo {title} {Stacking-driven gap formation in layered
  {1\emph{T}-TaS$_{2}$}},\ }\href@noop {} {\bibfield  {journal} {\bibinfo
  {journal} {Phys. Rev. B}\ }\textbf {\bibinfo {volume} {98}},\ \bibinfo
  {pages} {195134} (\bibinfo {year} {2018})}\BibitemShut {NoStop}%
\bibitem [{\citenamefont {Lee}\ \emph {et~al.}(2019)\citenamefont {Lee},
  \citenamefont {Goh},\ and\ \citenamefont {Cho}}]{Lee2019}%
  \BibitemOpen
  \bibfield  {author} {\bibinfo {author} {\bibfnamefont {S.-H.}\ \bibnamefont
  {Lee}}, \bibinfo {author} {\bibfnamefont {J.~S.}\ \bibnamefont {Goh}},\ and\
  \bibinfo {author} {\bibfnamefont {D.}~\bibnamefont {Cho}},\ }\bibfield
  {title} {\bibinfo {title} {Origin of the insulating phase and first-order
  metal-insulator transition in {$1T\text{-}{\mathrm{TaS}}_{2}$}},\ }\href@noop
  {} {\bibfield  {journal} {\bibinfo  {journal} {Phys. Rev. Lett.}\ }\textbf
  {\bibinfo {volume} {122}},\ \bibinfo {pages} {106404} (\bibinfo {year}
  {2019})}\BibitemShut {NoStop}%
\bibitem [{Lau(2015)}]{Laulhe2015}%
  \BibitemOpen
  \bibfield  {title} {\bibinfo {title} {{X}-ray study of femtosecond structural
  dynamics in the {2D} charge density wave compound
  \mbox{{1\emph{T}-TaS$_{2}$}}},\ }\href@noop {} {\bibfield  {journal}
  {\bibinfo  {journal} {Phys. B: Condens. Matter}\ }\textbf {\bibinfo {volume}
  {460}},\ \bibinfo {pages} {100} (\bibinfo {year} {2015})}\BibitemShut
  {NoStop}%
\bibitem [{\citenamefont {Hendricks}\ and\ \citenamefont
  {Teller}(1942)}]{Hendricks1942}%
  \BibitemOpen
  \bibfield  {author} {\bibinfo {author} {\bibfnamefont {S.~B.}\ \bibnamefont
  {Hendricks}}\ and\ \bibinfo {author} {\bibfnamefont {E.}~\bibnamefont
  {Teller}},\ }\bibfield  {title} {\bibinfo {title} {X‐ray interference in
  partially ordered layer lattices},\ }\href@noop {} {\bibfield  {journal}
  {\bibinfo  {journal} {J. Chem. Phys.}\ }\textbf {\bibinfo {volume} {10}},\
  \bibinfo {pages} {147} (\bibinfo {year} {1942})}\BibitemShut {NoStop}%
\bibitem [{\citenamefont {Treacy}\ \emph {et~al.}(1991)\citenamefont {Treacy},
  \citenamefont {Newsam},\ and\ \citenamefont {Deem}}]{Treacy1991}%
  \BibitemOpen
  \bibfield  {author} {\bibinfo {author} {\bibfnamefont {M.~M.~J.}\
  \bibnamefont {Treacy}}, \bibinfo {author} {\bibfnamefont {J.~M.}\
  \bibnamefont {Newsam}},\ and\ \bibinfo {author} {\bibfnamefont {M.~W.}\
  \bibnamefont {Deem}},\ }\bibfield  {title} {\bibinfo {title} {A~general
  recursion method for calculating diffracted intensities from crystals
  containing planar faults},\ }\href@noop {} {\bibfield  {journal} {\bibinfo
  {journal} {Proc. R. Soc. Lond. A}\ }\textbf {\bibinfo {volume} {433}},\
  \bibinfo {pages} {499} (\bibinfo {year} {1991})}\BibitemShut {NoStop}%
\bibitem [{\citenamefont {Nicholson}\ \emph {et~al.}(2024)\citenamefont
  {Nicholson} \emph {et~al.}}]{Nicolson2024}%
  \BibitemOpen
  \bibfield  {author} {\bibinfo {author} {\bibfnamefont {C.~W.}\ \bibnamefont
  {Nicholson}} \emph {et~al.},\ }\bibfield  {title} {\bibinfo {title} {Gap
  collapse and flat band induced by uniaxial strain in {1\emph{T}-TaS$_{2}$}},\
  }\href@noop {} {\bibfield  {journal} {\bibinfo  {journal} {Phys. Rev. B}\
  }\textbf {\bibinfo {volume} {109}},\ \bibinfo {pages} {035167} (\bibinfo
  {year} {2024})}\BibitemShut {NoStop}%
\bibitem [{\citenamefont {Yu}\ \emph {et~al.}(2017)\citenamefont {Yu} \emph
  {et~al.}}]{LongYu2017}%
  \BibitemOpen
  \bibfield  {author} {\bibinfo {author} {\bibfnamefont {X.-L.}\ \bibnamefont
  {Yu}} \emph {et~al.},\ }\bibfield  {title} {\bibinfo {title} {{Electronic
  correlation effects and orbital density wave in the layered compound
  $1T$-TaS$_2$}},\ }\href@noop {} {\bibfield  {journal} {\bibinfo  {journal}
  {Phys.~Rev. B}\ } (\bibinfo {year} {2017})}\BibitemShut {NoStop}%
\bibitem [{\citenamefont {Petocchi}\ \emph {et~al.}(2022)\citenamefont
  {Petocchi} \emph {et~al.}}]{Petocchi2022}%
  \BibitemOpen
  \bibfield  {author} {\bibinfo {author} {\bibfnamefont {F.}~\bibnamefont
  {Petocchi}} \emph {et~al.},\ }\bibfield  {title} {\bibinfo {title} {Mott
  versus hybridization gap in the low-temperature phase of {$1T$-TaS$_2$}},\
  }\href@noop {} {\bibfield  {journal} {\bibinfo  {journal} {Phys. Rev. Lett.}\
  }\textbf {\bibinfo {volume} {129}},\ \bibinfo {pages} {016402} (\bibinfo
  {year} {2022})}\BibitemShut {NoStop}%
\bibitem [{\citenamefont {de~la Torre}\ \emph {et~al.}()\citenamefont {de~la
  Torre} \emph {et~al.}}]{delaTorre2025}%
  \BibitemOpen
  \bibfield  {author} {\bibinfo {author} {\bibfnamefont {A.}~\bibnamefont
  {de~la Torre}} \emph {et~al.},\ }\href@noop {} {\bibinfo {title} {Dynamic
  phase transition into a mixed-{CDW} state in {$1T$-TaS$_2$} via a thermal
  quench}},\ \Eprint {https://arxiv.org/abs/2407.07953} {arXiv:2407.07953}
  \BibitemShut {NoStop}%
\bibitem [{\citenamefont {Bozin}\ \emph {et~al.}(2023)\citenamefont {Bozin}
  \emph {et~al.}}]{Bozin2023}%
  \BibitemOpen
  \bibfield  {author} {\bibinfo {author} {\bibfnamefont {E.~S.}\ \bibnamefont
  {Bozin}} \emph {et~al.},\ }\bibfield  {title} {\bibinfo {title}
  {Crystallization of polarons through charge and spin ordering transitions in
  {$1T$-TaS$_2$}},\ }\href@noop {} {\bibfield  {journal} {\bibinfo  {journal}
  {Nat. Commun.}\ }\textbf {\bibinfo {volume} {14}},\ \bibinfo {pages} {7055}
  (\bibinfo {year} {2023})}\BibitemShut {NoStop}%
\bibitem [{\citenamefont {Ren}\ \emph {et~al.}(2022)\citenamefont {Ren} \emph
  {et~al.}}]{Ren2022}%
  \BibitemOpen
  \bibfield  {author} {\bibinfo {author} {\bibfnamefont {Z.}~\bibnamefont
  {Ren}} \emph {et~al.},\ }\bibfield  {title} {\bibinfo {title} {Plethora of
  tunable {W}eyl fermions in kagome magnet {Fe$_3$Sn$_2$} thin films},\
  }\href@noop {} {\bibfield  {journal} {\bibinfo  {journal} {npj Quantum
  Mater.}\ }\textbf {\bibinfo {volume} {7}},\ \bibinfo {pages} {109} (\bibinfo
  {year} {2022})}\BibitemShut {NoStop}%
\bibitem [{\citenamefont {Ekahana}\ \emph {et~al.}(2024)\citenamefont {Ekahana}
  \emph {et~al.}}]{Ekahana2024}%
  \BibitemOpen
  \bibfield  {author} {\bibinfo {author} {\bibfnamefont {S.~A.}\ \bibnamefont
  {Ekahana}} \emph {et~al.},\ }\bibfield  {title} {\bibinfo {title} {Anomalous
  electrons in a metallic kagome ferromagnet},\ }\href@noop {} {\bibfield
  {journal} {\bibinfo  {journal} {Nature}\ }\textbf {\bibinfo {volume} {627}},\
  \bibinfo {pages} {67} (\bibinfo {year} {2024})}\BibitemShut {NoStop}%
\bibitem [{\citenamefont {Venturini}\ \emph {et~al.}()\citenamefont {Venturini}
  \emph {et~al.}}]{Venturini2024}%
  \BibitemOpen
  \bibfield  {author} {\bibinfo {author} {\bibfnamefont {R.}~\bibnamefont
  {Venturini}} \emph {et~al.},\ }\href@noop {} {\bibinfo {title} {Electrically
  driven non-volatile resistance switching between charge density wave states
  at room temperature}},\ \Eprint {https://arxiv.org/abs/2412.13094}
  {arXiv:2412.13094} \BibitemShut {NoStop}%
\bibitem [{\citenamefont {Wu}\ and\ \citenamefont {Li}(2021)}]{Wu2021}%
  \BibitemOpen
  \bibfield  {author} {\bibinfo {author} {\bibfnamefont {M.}~\bibnamefont
  {Wu}}\ and\ \bibinfo {author} {\bibfnamefont {J.}~\bibnamefont {Li}},\
  }\bibfield  {title} {\bibinfo {title} {Sliding ferroelectricity in {2D van
  der Waals materials: R}elated physics and future opportunities},\ }\href@noop
  {} {\bibfield  {journal} {\bibinfo  {journal} {Proc. Natl. Acad. Sci. USA}\
  }\textbf {\bibinfo {volume} {118}},\ \bibinfo {pages} {e2115703118} (\bibinfo
  {year} {2021})}\BibitemShut {NoStop}%
\bibitem [{\citenamefont {Bian}\ \emph {et~al.}(2024)\citenamefont {Bian} \emph
  {et~al.}}]{Bian2024}%
  \BibitemOpen
  \bibfield  {author} {\bibinfo {author} {\bibfnamefont {R.}~\bibnamefont
  {Bian}} \emph {et~al.},\ }\bibfield  {title} {\bibinfo {title} {Developing
  fatigue-resistant ferroelectrics using interlayer sliding switching},\
  }\href@noop {} {\bibfield  {journal} {\bibinfo  {journal} {Science}\ }\textbf
  {\bibinfo {volume} {385}},\ \bibinfo {pages} {57} (\bibinfo {year}
  {2024})}\BibitemShut {NoStop}%
\bibitem [{\citenamefont {Yasuda}\ \emph {et~al.}(2024)\citenamefont {Yasuda}
  \emph {et~al.}}]{Yasuda2024}%
  \BibitemOpen
  \bibfield  {author} {\bibinfo {author} {\bibfnamefont {K.}~\bibnamefont
  {Yasuda}} \emph {et~al.},\ }\bibfield  {title} {\bibinfo {title} {Ultrafast
  high-endurance memory based on sliding ferroelectrics},\ }\href@noop {}
  {\bibfield  {journal} {\bibinfo  {journal} {Science}\ }\textbf {\bibinfo
  {volume} {385}},\ \bibinfo {pages} {53} (\bibinfo {year} {2024})}\BibitemShut
  {NoStop}%
\bibitem [{\citenamefont {Vizner~Stern}\ \emph {et~al.}(2025)\citenamefont
  {Vizner~Stern}, \citenamefont {Salleh~Atri},\ and\ \citenamefont
  {Ben~Shalom}}]{Stern2025}%
  \BibitemOpen
  \bibfield  {author} {\bibinfo {author} {\bibfnamefont {M.}~\bibnamefont
  {Vizner~Stern}}, \bibinfo {author} {\bibfnamefont {S.}~\bibnamefont
  {Salleh~Atri}},\ and\ \bibinfo {author} {\bibfnamefont {M.}~\bibnamefont
  {Ben~Shalom}},\ }\bibfield  {title} {\bibinfo {title} {Sliding van der
  {W}aals polytypes},\ }\href@noop {} {\bibfield  {journal} {\bibinfo
  {journal} {Nat. Rev. Phys.}\ }\textbf {\bibinfo {volume} {7}},\ \bibinfo
  {pages} {50} (\bibinfo {year} {2025})}\BibitemShut {NoStop}%
\bibitem [{\citenamefont {Walker}\ and\ \citenamefont
  {Withers}(1983)}]{Walker1983}%
  \BibitemOpen
  \bibfield  {author} {\bibinfo {author} {\bibfnamefont {M.~B.}\ \bibnamefont
  {Walker}}\ and\ \bibinfo {author} {\bibfnamefont {R.~L.}\ \bibnamefont
  {Withers}},\ }\bibfield  {title} {\bibinfo {title} {Stacking of
  charge-density waves in {$1T$} transition-metal dichalcogenides},\
  }\href@noop {} {\bibfield  {journal} {\bibinfo  {journal} {Phys. Rev. B}\
  }\textbf {\bibinfo {volume} {28}},\ \bibinfo {pages} {2766} (\bibinfo {year}
  {1983})}\BibitemShut {NoStop}%
\bibitem [{\citenamefont {Brouwer}\ and\ \citenamefont
  {Jellinek}(1980)}]{Brouwer1980}%
  \BibitemOpen
  \bibfield  {author} {\bibinfo {author} {\bibfnamefont {R.}~\bibnamefont
  {Brouwer}}\ and\ \bibinfo {author} {\bibfnamefont {F.}~\bibnamefont
  {Jellinek}},\ }\bibfield  {title} {\bibinfo {title} {The low-temperature
  superstructures of {1\emph{T}-TaSe$_{2}$} and {2\emph{H}-TaSe$_{2}$}},\
  }\href@noop {} {\bibfield  {journal} {\bibinfo  {journal} {Physica B+C}\
  }\textbf {\bibinfo {volume} {99}},\ \bibinfo {pages} {51} (\bibinfo {year}
  {1980})}\BibitemShut {NoStop}%
\bibitem [{\citenamefont {Zong}\ \emph {et~al.}(2018)\citenamefont {Zong} \emph
  {et~al.}}]{Zong2018}%
  \BibitemOpen
  \bibfield  {author} {\bibinfo {author} {\bibfnamefont {A.}~\bibnamefont
  {Zong}} \emph {et~al.},\ }\bibfield  {title} {\bibinfo {title} {Ultrafast
  manipulation of mirror domain walls in a charge density wave},\ }\href@noop
  {} {\bibfield  {journal} {\bibinfo  {journal} {Sci.~Adv.}\ }\textbf {\bibinfo
  {volume} {4}},\ \bibinfo {pages} {eaau5501} (\bibinfo {year}
  {2018})}\BibitemShut {NoStop}%
\bibitem [{\citenamefont {Frenkel}\ \emph {et~al.}(2017)\citenamefont
  {Frenkel}, \citenamefont {Schrenk},\ and\ \citenamefont
  {Martiniani}}]{Frenkel2017}%
  \BibitemOpen
  \bibfield  {author} {\bibinfo {author} {\bibfnamefont {D.}~\bibnamefont
  {Frenkel}}, \bibinfo {author} {\bibfnamefont {K.~J.}\ \bibnamefont
  {Schrenk}},\ and\ \bibinfo {author} {\bibfnamefont {S.}~\bibnamefont
  {Martiniani}},\ }\bibfield  {title} {\bibinfo {title} {Monte {C}arlo sampling
  for stochastic weight functions},\ }\href@noop {} {\bibfield  {journal}
  {\bibinfo  {journal} {Proc. Natl. Acad. Sci. USA}\ }\textbf {\bibinfo
  {volume} {114}},\ \bibinfo {pages} {6924} (\bibinfo {year}
  {2017})}\BibitemShut {NoStop}%
\end{thebibliography}%

\section{End Matter}

\subsection{Probing Stacking Dynamics with Coherent X-rays}

\begin{figure}[b]
	\includegraphics[width=1\columnwidth]{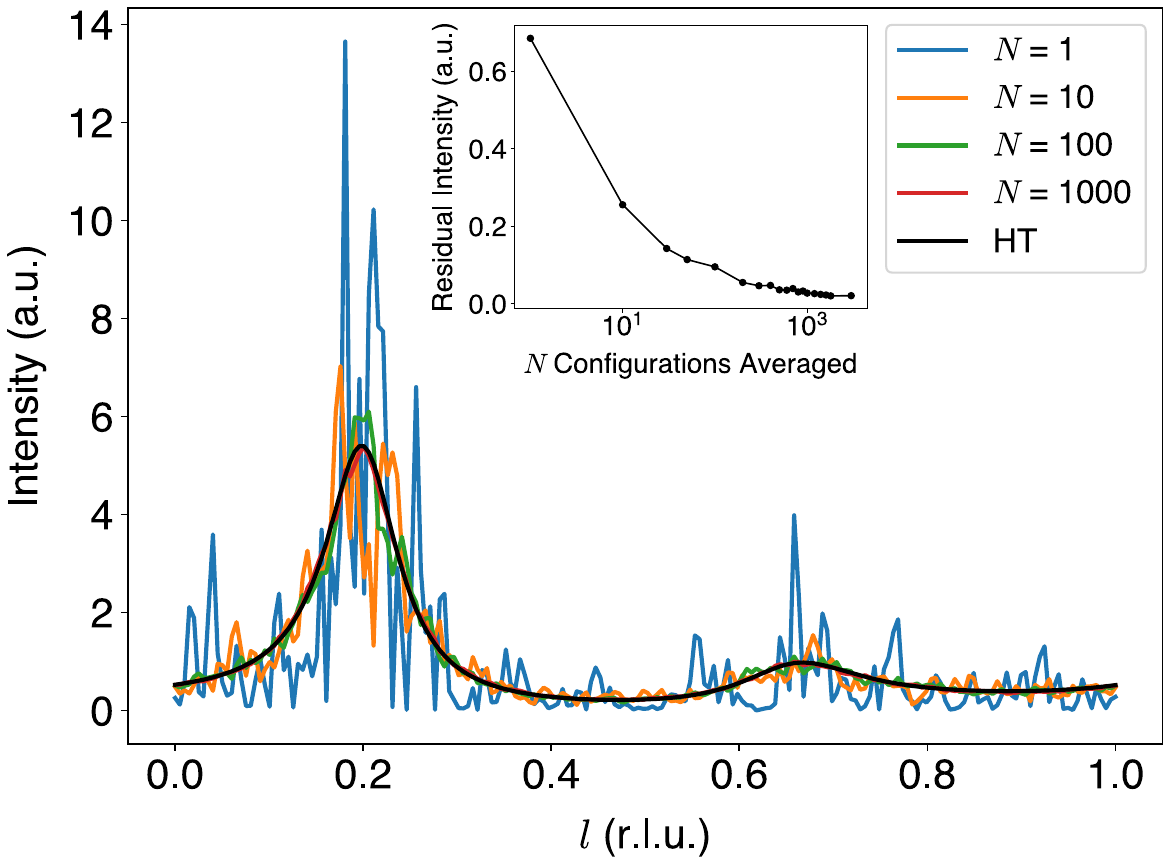}
	\caption{\label{Fig5}
	Simulated XRD signal along the (0.308, -0.077, $l$) direction in reciprocal space of a 100-layer \mbox{1\emph{T}-TaS$_{2}$} stacking configuration with a dimer-to-monolayer ratio of 2:1. The analytical HT solution (black) is overlaid with the results from averaging $N$ configurations generated by the MC method. The inset shows the intensity differences between the HT result and MC signals from $N$~averaged configurations.}
\end{figure}

Beyond replicating the XRD signals originating from the static CCDW stacking structure, the numerical MC and analytical HT formalisms have implications for the ability to directly visualize stacking dynamics in a coherent X-ray scattering experiment. Figure~\ref{Fig5} shows the intensity of a CCDW diffraction peak generated via the HT approach, and averages of $N$ configurations produced by the stochastic MC method. Averaging $N \to \infty$ configurations from MC reproduces the HT solution, demonstrating their equivalence. This is quantified in the inset of Fig.~\ref{Fig5}. For the case $N = 1$, the residual intensity between the two formalisms shows a large variance with points of zero intensity. Here, the stacking structure effectively acts as a diffraction grating for a coherent X-ray beam with a small spot size that produces a speckle pattern. The MC method can therefore be used to simulate the CCDW diffraction pattern generated by partially and fully coherent X-rays,  where stacking dynamics manifests as a fluctuating speckle pattern.

\subsection{Stacking Order of the Nearly-Commensurate and Hidden CDW States of 1\textit{T}-TaS$_2$}

\begin{figure}[!b]
	\includegraphics[width=1\columnwidth]{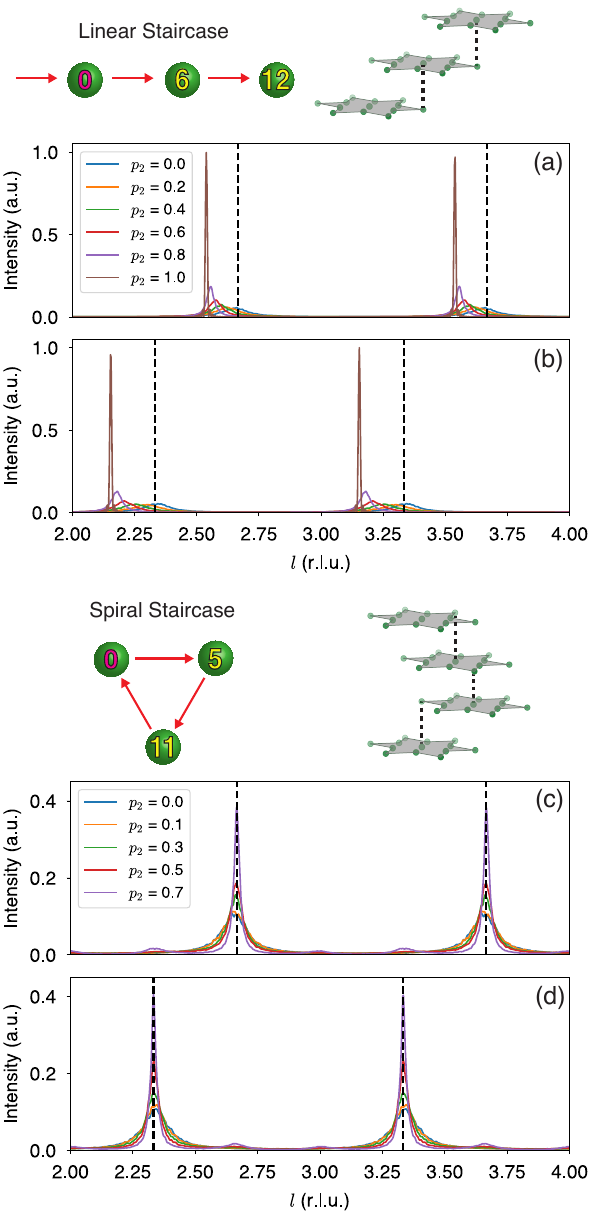}
	\caption{\label{Fig6}
		Linear and a spiral staircase stacking are schematically shown on the top and middle, respectively. Ta atoms are numbered as in Fig.~1 and refer to the stacking order. The XRD intensity are shown along \mbox{(a,c) (0.077, 0.231, $l$)} and (b,d) (0.308, $-0.077$, $l$), corresponding to the two symmetry equivalent groups described in Fig. 1, and as a function of the order $p_{2}$. The vertical dotted lines refer to the $l$ positions where the NCCDW and HCDW peaks appear in experiments.}
\end{figure}

Disordered stacking is simulated by assuming that the stacking probabilities within a symmetry group are equal. For example, in the \emph{T$_{c1}$} group, the stackings of Ta atoms \{7, 8, 11\} above the central Ta atom are all equally likely. In this manner, we can exhaust all possible disordered stacking distributions. Ordered stacking structures and the degree of order can also be simulated with the HT and MC formalisms. Several ordered stacking structures in \mbox{1\emph{T}-TaS$_{2}$} have previously been proposed \cite{Walker1983}. Figure \ref{Fig6} shows two possibilities, corresponding to a linear and spiral staircase \emph{T$_{c}$} stacking order. A thorough description of the mathematical formalism for simulating the stacking order is presented in the Supplementary Material.

The simulated stacking can be compared to the experimental Bragg peak positions of the nearly-commensurate~(NCCDW) and the nonthermal hidden~(HCDW) charge-density wave phases of \mbox{1\emph{T}-TaS$_{2}$}~\cite{Stahl2020}. Figures~\ref{Fig6}(a,b) show the out-of-plane XRD intensity from a linear staircase stacking, where the vertical dotted lines denote the $l$ = $\pm \frac{1}{3}$ locations of the HCDW and NCCDW peaks. Due to the S~atoms that break the six-fold symmetry (Fig. 1) two in-plane CDW coordinates are shown to capture the symmetry breaking. Clearly, this stacking structure does not describe those CDW phases, while the spiral staircase stacking structure in Fig.~\ref{Fig6}(c,d) yields diffraction patterns that align with the experiments. Here, $p_{2}$ describes the degree of ordered stacking, corresponding to the length scale before a stacking fault is introduced. Although the interlayer correlation length of the NCCDW and HCDW phases is larger than that of the CCDW phase~\cite{Stahl2020}, there is no experimental evidence for additional out-of-plane superlattice signals, \textit{e.g.} originating from dimers, in those phases. However, additional peaks appear in Fig.~\ref{Fig6}(c,d) when $p_{2} \gtrsim$ 0.7. Based on the peak location and intensity distribution along the out-of-plane $l$ direction, we therefore assign a spiral staircase ordered stacking with $p_{2} \approx$ 0.6 for the NCCDW and HCDW states.


\subsection{Out-of-Plane Dimer Contraction}

\begin{figure}[!ht]
	\includegraphics[width=1\columnwidth]{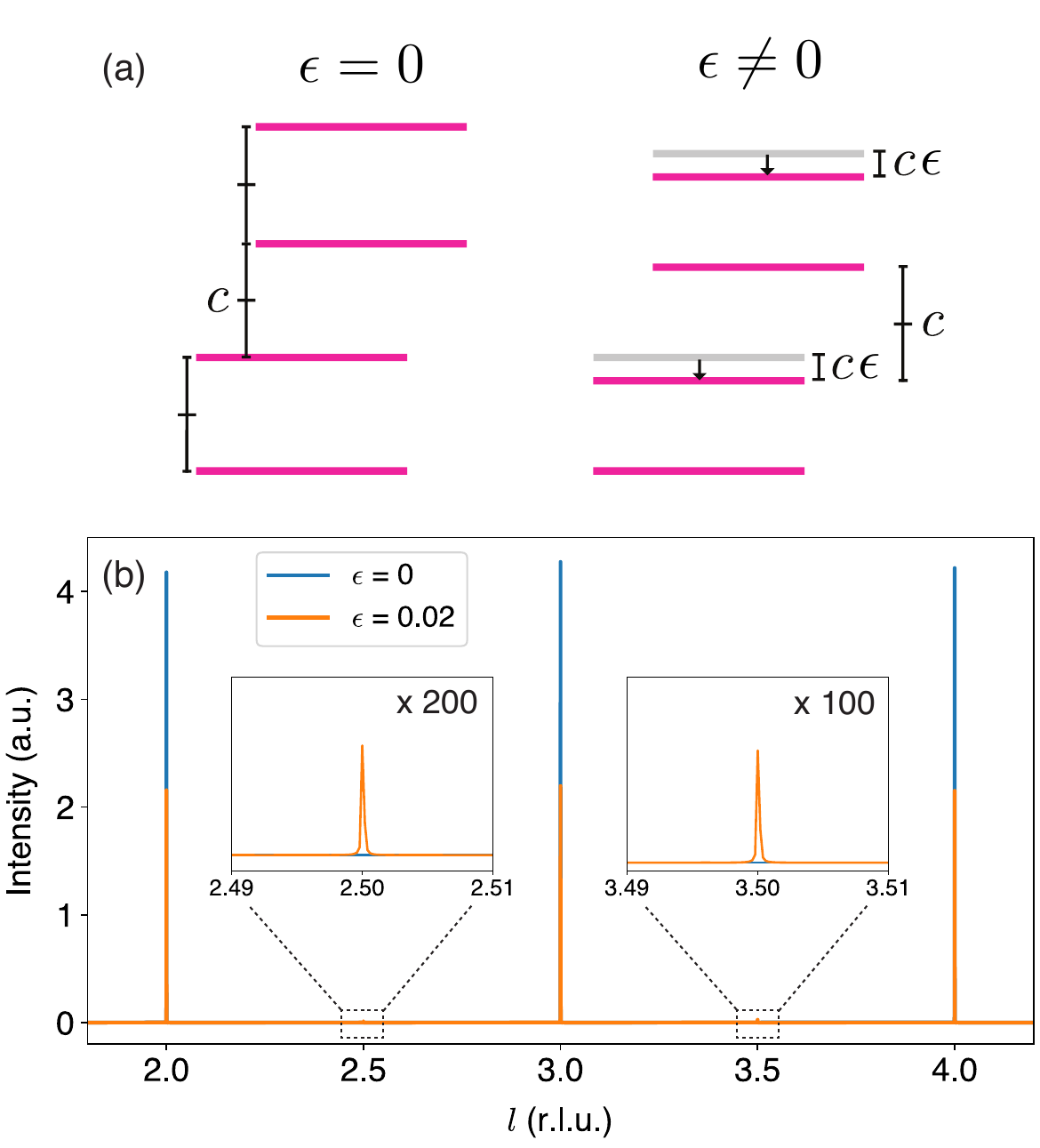}
	\caption{\label{Fig7}{(a) Schematic of two \emph{T$_{c}$}-stacked dimers with identical \emph{intra} and \emph{inter} dimer distances \mbox{(left, $\epsilon = 0$)} and an \emph{intra} dimer contraction (right, $\epsilon \neq 0$}). (b)~Simulated (0, 0, $l$) lattice peak intensity, where weak half-ordered superlattice signals appear in the contracted case \mbox{($\epsilon = 0.02$)}.} 
\end{figure}

XRD directly reveals the presence of dimers in the CCDW phase via half-ordered CDW peaks \cite{Tanda1984, Stahl2020}. In addition, an overall out-of-plane lattice contraction, estimated to be 0.43\%, is observed proceeding from the NCCDW phase to the low-temperature CCDW state~\cite{Wang2020}. What remains to be investigated is whether the \emph{intra} dimer distance (\emph{T$_{a}$}) and the distance between dimers/monolayers (\emph{T$_{c}$}) are the same or not. We use the HT formalism to look into this question by introducing the parameter $\epsilon$, defined as the fraction of the $c$ lattice constant that a dimer contracts in the CCDW phase. As shown in Fig. \ref{Fig7}(a), the separation between dimerized layers changes from $c$ to $(1-\epsilon)c$, whereas the distance between the two dimerized pairs remains $c$. This parametrization allows for a different lattice spacing between dimers, pairs of dimers, as well as monolayers and dimers. A detailed calculation is presented in the Supplementary Material, but setting $\epsilon = 0.02$ and calculating lattice diffraction peaks along the out-of-plane (0, 0, $l$) direction, we find half-ordered lattice diffraction peaks, in line with XRD experiments \cite{Wang2020, Burri2024}. The relative intensities of the half-ordered peaks from our simulations are about a magnitude larger than seen experimentally, implying $\epsilon \ll 0.02$.

\clearpage
\FloatBarrier

\setcounter{section}{0}
\renewcommand{\thesection}{S\arabic{section}}
\setcounter{figure}{0}
\renewcommand{\thefigure}{S\arabic{figure}}
\setcounter{table}{0}
\renewcommand{\thetable}{S\arabic{table}}
\setcounter{equation}{0}
\renewcommand{\theequation}{S\arabic{equation}}

\renewcommand{\thefootnote}{\fnsymbol{footnote}}

\begin{center}
    \Large \textbf{Supplementary Information} \\

\end{center}
\section{Structure Factor Calculations}\label{sec:structure}

As described in the main text, the structure factor for a layered material, assuming no in-plane disorder, can be written as

\begin{equation} \label{Seq1}
    F(\vec{Q}) = F_{\rm planar}(\vec{Q}) \sum_{n=1}^N e^{-i\vec{Q} \cdot \Delta \vec{r}_n}.
\end{equation}

The atomic positions of the undistorted lattice of 1\textit{T}-TaS$_2$ are set based on the lattice constants \mbox{$a=b$~=~0.3382~nm} and $c$ = 0.5892 nm, while the distortion of the in-plane polarons was taken from~\cite{Brouwer1980}. This reproduces the in-plane charge-density wave~(CDW) diffraction peaks that are consistent with the ($h,~k$) coordinates seen experimentally in \mbox{X-ray} diffraction~(XRD)~\cite{Laulhe2015, Stahl2020}. We note that certain experimental contributions, such as the Debye-Waller factor, the Lorentz polarization factor, and the photon flux, all of which affect the absolute intensities of diffraction peaks in real experiments, were not taken into account in these simulations. Nevertheless, they show relative changes in the peak positions, intensities, and widths in reciprocal space due to variation of the stacking structure. 

The ordered in-plane structure yields bright Bragg peaks localized in the ($h$,~$k$) plane,  while the disordered out-of-plane structure produces long diffuse peaks along $l$-direction in the commensurate CDW (CCDW) phase. The combination of these two projections results in pillar-like peaks in three-dimensional reciprocal space. We are primarily interested in the interactions that produce the disordered structure and not a specific stacking configuration. With this in mind, we can model the diffracted intensity of a disordered crystal as the average of the intensity over all stacking configurations weighted by their probability,

\begin{equation} \label{Seq2}
\langle I (\vec{Q}) \rangle_{\{X\}} = \sum_{x \in \{X\}} P(x) I(\vec{Q})_{x},
\end{equation}

\noindent where $\{X\}$ represents the set of possible configurations and $P(x)$ is the probability of a certain configuration $x$, which by construction must be normalized to $1$. The intensity of a configuration $I(\vec{Q})_x$ can be computed with Eq.~\eqref{Seq1}. For a crystal with $N$ layers and $K$~different types of layers, the number of total configurations is $K^N$. For the CCDW state of 1\emph{T}-TaS\textsubscript{2}, the number of possible stacking layers in the simplest case is $K = 13$ (assuming Ta atoms stack above each other). With a conservative number of layers of $N = 80$ for a bulk-like flake,  the number of configurations is $\approx1.3 \times 10^{89}$, meaning that direct evaluation of the sum in Eq.~(\ref{Seq2}) is impractical. Furthermore, we are interested in the \emph{distribution} of stacking configurations, and not necessarily the \emph{exact} stacking configuration. Two separate methods, a numerical Monte Carlo (MC) approach and an analytical method based on the Hendricks-Teller (HT) formalism of stacking faults, are used to account for the distribution of stacking configurations. 

\section{Symmetry of Polaron Stars}

Here we show how to parameterize the nearest neighbor stacking options for the polaron stars in 1\emph{T}-TaS\textsubscript{2} using the smallest number of variables given by symmetry constraints. The space group of a single layer of 1\emph{T}-TaS\textsubscript{2} in the CCDW state is \textit{P}$\overline{3}$ (number 147). This group is symmorphic and contains the point group $C_{3i} = S_6$ that consists of rotations by 120~degrees and inversion about the central Ta atom. 

\begin{figure*}[!ht]
    \centering
	\includegraphics[width=.75\textwidth]{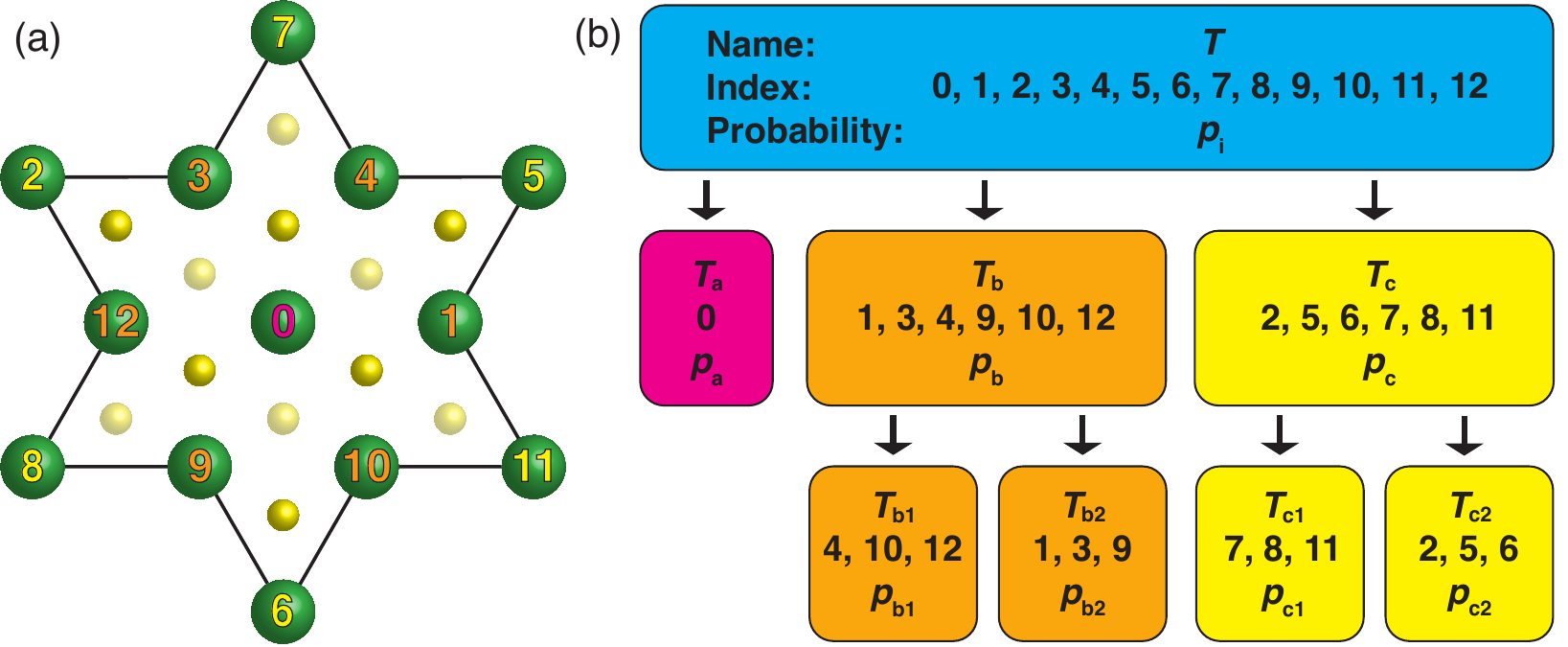}
	\caption{\label{SFig1}
	(a) A polaron star composed of Ta atoms (green) and S~atoms~(yellow) above and below the plane of Ta atoms. (b)~A flowchart shows a general stacking vector \textit{T} can be decomposed into three groups with different magnitudes: $T_a$, $T_b$, or $T_c$. These groups can be further separated into five symmetry equivalent vectors with their own distinct probabilities.}
\end{figure*}

To deal with the periodicity of the structure, an index convention for each Ta within a polaron is assigned. Starting with 0 for the central Ta atom, sequential indices are assigned by translating by $\vec{a}$ (see Fig.~\ref{SFig1}). Once the translation hops off the boundary of the center polaron, we move back to the same Ta as on the outer star. This encodes the periodicity of the polaron structure and is unique to each chirality \cite{Zong2018}. In our simulations, though, the assumption of in-plane order means we also assume the system is of one chirality which is also what one typically finds in XRD. 

 We are primarily interested in invariances of the interactions between the order parameter accounting for commensurate modulations in neighboring planes. In other words, which relative translations between the modulations of two layers are connected by symmetry. We can describe the phase of the CDW modulation by the position of the center Ta atom in each plane. By fixing the center Ta atom of the bottom plane at the origin, the relative phase can be defined by the position of the center Ta atom on the second layer $\vec{R}$, which we can label as an index $i \in [0,12]$ that we name \textit{T}. Three different lengths of stacking vectors can be identified: $T_a$, $T_b$, and $T_c$ defined by translation to the center Ta atom, as well as the inner and outer vertex of the polaron, respectively. Due to the translational symmetry of the polaron planes, any translation larger than $\vec{R}$ is equivalent to a $T_a$, $T_b$, or $T_c$ stacking. However, not all $T_b$ or $T_c$ stacking vectors are connected by symmetry. Only those that are rotated 120~degrees from each other are symmetrical due to the lack of six-fold rotational symmetry of the S atoms. Therefore, we can further separate $T_c$ stacking into $T_{c1}$ and $T_{c2}$ and the same for $T_b$ \cite{Walker1983}. The inversion symmetry of the point group is not relevant, because this would swap the top plane to below the origin plane that would produce the same stacking vector. A physical representation of this analysis is shown in Fig. \ref{SFig1}. This symmetry analysis is only valid for nearest neighbor (NN) stacking interactions. Next-nearest neighbor~(NNN) interactions can break the degeneracy of the symmetry equivalent stacking vectors, discussed in the next section.

\section{Hendricks-Teller Model}

A general theory involving probabilistic nearest neighbor matrices and different types of layers to solve Eq.~\eqref{Seq2} for a system of one-dimensional (1D) disorder was formulated by Hendricks and Teller in 1942 \cite{Hendricks1942}. An equivalent theory will be described below, formulated more recently by Treacy~\emph{et~al}.~\cite{Treacy1991}. It involves a recursive variation of the original HT~formalism and was used to study materials such as zeolytes and diamond-lonsdaleite, but can be equally applied to disordered quantum materials.

The probability of a general configuration of $n$ layer types can be written as follows 

\begin{equation} \label{probability matrix}
P(x = ijkl...mn) = g_i A_{ij} A_{jk} A_{kl} ... A_{mn},
\end{equation}

\noindent where $g_i$ is a vector containing the \emph{a priori} probability for layer type $i$ to exist. The stacking matrix $A_{ij}$ gives the probability that layer type $i$ is followed by a $j$-type layer. This relies on introducing different types of layers labeled by $i$ and breaking up the probability for a stack of layers to exist with the product of probabilities of layer type $i$ to be next to $j$-type layer. By defining these objects as probabilities, the conditions that $\sum_i g_i = 1$ and $\sum_j A_{ij} = 1$ follow. The vector $g_i$ can be found as the stationary eigenvector of $A_{ij}$. In the interest of computing the intensity for a certain configuration, we can write the total scattered wave from $N$ layers as 

\begin{equation}
\Phi_{x = ijk...}(\vec{Q})
 = F_i + F_j e^{i\vec{Q} \cdot \vec{R}_{ij}} + F_k e^{i\vec{Q} \cdot (\vec{R}_{ij} + \vec{R}_{jk})} + ...,
\end{equation}

\noindent which is just the sum of the scattered waves from each plane including a weighting by the phase factor acquired from its position in the stack. Here, $\vec{R}_{ij}$ is the stacking vector between layers of type $i$ and $j$. We can then write the intensity as \mbox{$I(x = ijk...) = \Phi^*_{x = ijk...}(\vec{Q}) \Phi_{x = ijk...}(\vec{Q})$}. Combining this with Eq.~\eqref{probability matrix}, we can rewrite Eq.~\eqref{Seq2} as 

\begin{equation} \label{Hendricks-Teller-Setup}
    I(\vec{Q}) = \sum_{x = ijkl...} g_i A_{ij} A_{jk} A_{kl} ...  |\Phi_{x = ijk...}(\vec{Q})|^2.
\end{equation}

\noindent With clever manipulations of Eq.~\eqref{Hendricks-Teller-Setup}, it can be evaluated as a geometric series to yield the following result

\begin{equation} \label{Hendriks-Teller Result}
I(\vec{Q})/N = \vec{G}^* \cdot \vec{\psi} +\vec{G} \cdot \vec{\psi}^* - \vec{G}^* \cdot \vec{F},    
\end{equation}

\noindent where the vector $G_i = g_i F_i$ and $\psi$ is 

$$\vec{\psi} = \frac{(I-T)^{-1}}{N}\left( (N+1)I - (I - T)^{-1}(I-T^{N+1})\right) \vec{F}.$$

Here, $T$ is a complex transition probability matrix $T_{ij}(\vec{Q}) = A_{ij} e^{i\vec{Q} \cdot \vec{R}_{ij}}$. With Eq.~\eqref{Hendriks-Teller Result}, the scattering intensity for a crystal with 1D disorder can be computed. For simple systems with a small number of layers or a sparse matrix $A_{ij}$, Eq.~\eqref{Hendriks-Teller Result} can be evaluated by hand. For more complicated systems and a dense stacking matrix, exact numerical evaluation is preferred. Equation~\eqref{Hendriks-Teller Result} can be used to evaluate the intensity for all systems with 1D disorder and follows from Eq.~\eqref{Seq2} without assumptions. Thus, the result is applicable to many scenarios where disorder in 1D is present. It may seem that this model assumes only NN interactions, but in fact a system with NNN interactions can be mapped to an identical system with NN interactions. This is analogous to an $N$-order Monte Carlo chain that can be transformed into a first-order one. It quickly becomes computationally expensive with increasing number of nearby layers that are considered, but is still feasible for second NN~interactions.

\section{HT Model Applied to {1\emph{T}-T\lowercase{a}S$_{2}$}}

\subsection{Monolayer XRD Intensity Simulation}

In this section, we apply Eq.~\eqref{Hendriks-Teller Result} to 1\emph{T}-TaS\textsubscript{2} and consider the simplest form of monolayer 1\emph{T}-TaS\textsubscript{2} planes with NN interactions. Although we assume that all planes are identical, for the purpose of the computation, 13 different layer types are necessary for the 13~different stacking vectors. As described previously, the 13 stacking vectors can be grouped into five symmetry-equivalent groups labeled $T_a, T_{b1}, T_{b2}, T_{c1},$ and $T_{c2}$ that have a stacking probability given by the symmetry of the system, labeled as $p_a, p_b p_{b1}, p_b p_{b2}, p_c p_{c1},$ and $p_c p_{c2}$. $p_{a/b/c}$ describes the probability for $T_{a/b/c}$ stacking,  while $p_{c1/c2}$ is the conditional probability to stack to one of the symmetry inequivalent groups given $c$ or $b$ stacking. Based on these definitions and the condition of probability normalization, the following must hold 

\begin{equation} 
p_a + p_b + p_c = 1, \hspace{.45 cm} p_{b1} + p_{b2} = 1, \hspace{.45 cm} p_{c1} + p_{c2} = 1.
\end{equation}

With seven free variables and three conditions, this model of the NN stacking has four degrees of freedom. With these definitions, a succinct way to compute the stacking probability matrix $A_{ij}^{(1)}$ for NN interactions is

\begin{equation} \label{A_matrix}
    A_{ij}^{(1)} = \sum_{i=n}^{13} p_n P_n,
\end{equation}

\noindent where $p_n$ is the probability to go from position $0$ to $n$ (see Fig. \ref{SFig1}) and $P_n$ is the 13 x 13 permutation matrix which has ones where $j = i+n$. It can also be seen that $P_n = (P_1)^n$. Using this equation, the transition probability matrix can be transformed into Eq.~(S9) where the layers are labeled according to Fig. \ref{SFig1}. The $3 \times 13 \times 13$ matrix $R_{ij}$ containing the stacking vectors can be computed with a method similar to the one shown in Eq.~\eqref{A_matrix}. The symmetry of the system is encoded in $A_{ij}$. This can be seen from $R A R^{-1} = A$, where $R \in G_R = \{e, R_{2\pi/3}, R_{2\pi/3}\}$ and $G_R$ is the representation of the rotation subgroup of the point group. The operation $R_\theta$ permutes each atom to its position, rotated by $\theta$ around the center Ta atom. In general, $A_{ji} \neq A_{ij}$. The physical meaning of this inequality is that stacking from layers of type $i$ to $j$ does not have the same probability as stacking from $j$~to~$i$. This is a consequence of the definition $p_{c1} \neq p_{c2}$, and  directly linked to the lack of $xy$ mirror plane symmetry, as can also been seen from the scattering data: for a CDW peak at $l = n + 0.2$, where $n$ is an integer, there is no peak at the same in-plane position at $l = n-0.2$, or in general $I(h,k, l) \neq I(h,k, -l)$. 

\begin{figure*}[!ht]
\centering
\small
\begin{equation} \label{stacking probability matrix}
A_{ij}^{(1)} = \frac{1}{3}
\begin{bmatrix}
3p_a  & p_b p_{b2} & p_c p_{c1} & p_b p_{b2} & p_b p_{b1} & p_c p_{c1} & p_c p_{c1} & p_c p_{c2} & p_c p_{c2} & p_b p_{b2} & p_b p_{b1} & p_c p_{c2} & p_b p_{b1} \\
p_b p_{b1} & 3p_a  & p_b p_{b2} & p_c p_{c1} & p_b p_{b2} & p_b p_{b1} & p_c p_{c1} & p_c p_{c1} & p_c p_{c2} & p_c p_{c2} & p_b p_{b2} & p_b p_{b1} & p_c p_{c2} \\
p_c p_{c2} & p_b p_{b1} & 3p_a  & p_b p_{b2} & p_c p_{c1} & p_b p_{b2} & p_b p_{b1} & p_c p_{c1} & p_c p_{c1} & p_c p_{c2} & p_c p_{c2} & p_b p_{b2} & p_b p_{b1} \\
p_b p_{b1} & p_c p_{c2} & p_b p_{b1} & 3p_a  & p_b p_{b2} & p_c p_{c1} & p_b p_{b2} & p_b p_{b1} & p_c p_{c1} & p_c p_{c1} & p_c p_{c2} & p_c p_{c2} & p_b p_{b2} \\
p_b p_{b2} & p_b p_{b1} & p_c p_{c2} & p_b p_{b1} & 3p_a  & p_b p_{b2} & p_c p_{c1} & p_b p_{b2} & p_b p_{b1} & p_c p_{c1} & p_c p_{c1} & p_c p_{c2} & p_c p_{c2} \\
p_c p_{c2} & p_b p_{b2} & p_b p_{b1} & p_c p_{c2} & p_b p_{b1} & 3p_a  & p_b p_{b2} & p_c p_{c1} & p_b p_{b2} & p_b p_{b1} & p_c p_{c1} & p_c p_{c1} & p_c p_{c2} \\
p_c p_{c2} & p_c p_{c2} & p_b p_{b2} & p_b p_{b1} & p_c p_{c2} & p_b p_{b1} & 3p_a  & p_b p_{b2} & p_c p_{c1} & p_b p_{b2} & p_b p_{b1} & p_c p_{c1} & p_c p_{c1} \\
p_c p_{c1} & p_c p_{c2} & p_c p_{c2} & p_b p_{b2} & p_b p_{b1} & p_c p_{c2} & p_b p_{b1} & 3p_a  & p_b p_{b2} & p_c p_{c1} & p_b p_{b2} & p_b p_{b1} & p_c p_{c1} \\
p_c p_{c1} & p_c p_{c1} & p_c p_{c2} & p_c p_{c2} & p_b p_{b2} & p_b p_{b1} & p_c p_{c2} & p_b p_{b1} & 3p_a  & p_b p_{b2} & p_c p_{c1} & p_b p_{b2} & p_b p_{b1} \\
p_b p_{b1} & p_c p_{c1} & p_c p_{c1} & p_c p_{c2} & p_c p_{c2} & p_b p_{b2} & p_b p_{b1} & p_c p_{c2} & p_b p_{b1} & 3p_a  & p_b p_{b2} & p_c p_{c1} & p_b p_{b2} \\
p_b p_{b2} & p_b p_{b1} & p_c p_{c1} & p_c p_{c1} & p_c p_{c2} & p_c p_{c2} & p_b p_{b2} & p_b p_{b1} & p_c p_{c2} & p_b p_{b1} & 3p_a  & p_b p_{b2} & p_c p_{c1} \\
p_c p_{c1} & p_b p_{b2} & p_b p_{b1} & p_c p_{c1} & p_c p_{c1} & p_c p_{c2} & p_c p_{c2} & p_b p_{b2} & p_b p_{b1} & p_c p_{c2} & p_b p_{b1} & 3p_a  & p_b p_{b2} \\
p_b p_{b2} & p_c p_{c1} & p_b p_{b2} & p_b p_{b1} & p_c p_{c1} & p_c p_{c1} & p_c p_{c2} & p_c p_{c2} & p_b p_{b2} & p_b p_{b1} & p_c p_{c2} & p_b p_{b1} & 3p_a  \\
\end{bmatrix},
\end{equation}
\end{figure*}
\normalsize

\begin{figure} 
\begin{center}

\includegraphics[width=.95\columnwidth]{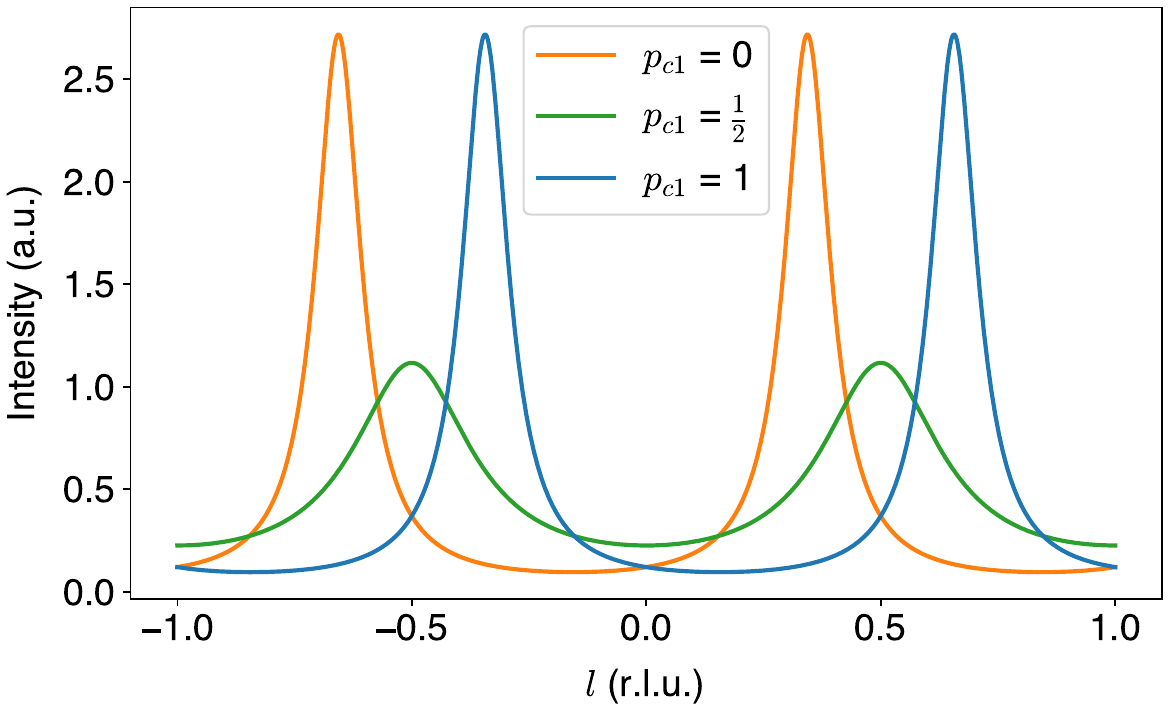} 
\caption{XRD intensity with only $T_c$ stacking but variable~$p_{c1}$. The characteristics of disordered stacking can be seen in the large peak width. Equation~\eqref{Hendriks-Teller Result} was evaluated for $l \in [-1,1]$ at the in-plane position ($h$ = $\frac{1}{13}$, $k$ = $\frac{3}{13}$) for three different $p_{c1}$, which has the effect of favoring different symmetry inequivalent $T_c$ stacking vectors.}
\label{monolayer_stacking}
\end{center}
\end{figure}

With $\vec{F} = 1$, $\vec{g} = \frac{1}{13}$, and because the planes have an identical structure, we can evaluate Eq.~\eqref{Hendriks-Teller Result}. Due to the dense $A_{ij}$ matrix, the inverse and matrix power are impractical to compute  analytically. But this task is straightforward  numerically, and just  requires that Eq.~\eqref{Hendriks-Teller Result} is evaluated for each value of $\vec{Q}$. Three intensity calculations are shown in Fig. $\ref{monolayer_stacking}$ with $h = \frac{1}{13}$, $k = \frac{3}{13}$, and $l \in [-1,1]$.

These three calculations all have only $T_c$ stacking but different dominant stacking vectors, following the convention of Fig. \ref{SFig1}. It can be seen that the calculation where $p_{c1} = \frac{1}{2}$ has reflection symmetry about the origin, $I(h,k,l) = I(h,k,-l)$, while the other two do not share this symmetry --- they transform onto each other with the $xy$ mirror plane reflection. It can also be seen that at $p_{c1} = 1$ or $p_{c1} = 0$, the peak center is at $l = \pm \frac13$. This aligns with the $l$ position of the peaks for 1\emph{T}-TaS\textsubscript{2} in the nearly-commensurate (NCCDW) and hidden~(HCDW) CDW states. However, the correlation lengths of those peaks are found to be much larger than the simulation in Fig. \ref{monolayer_stacking}, which is due to the presence of ordered stacking resulting from longer-ranged 
interactions between the layers.

\subsection{Dimer XRD Intensity Simulation}

The results of Ref.~\cite{Stahl2020} show a dimerized peak at an out-of-plane position that is \mbox{$\approx\frac{1}{2}$}~reciprocal lattice units~(r.l.u.) away from the main CDW peaks at \mbox{$l$ = $0.2$~r.l.u.}, suggesting the presence of dimerized layers. With the interest of simulating the CCDW state of 1\emph{T}-TaS\textsubscript{2}, dimerization must be considered as this is also the proposed mechanism for the band insulating nature of bulk 1\emph{T}-TaS\textsubscript{2}. Although in our current model $T_a$ stacking can be accounted for with a nonzero $p_a$, the second NN statistics of the dimerization interaction cannot. For example, for a system where dimerization is likely but $p_a$ is small, a layer that is already dimerized will be less likely to have a $T_a$ stacked layer above it than an unbound monolayer. This behavior could be added by including NNN interactions, but it turns out to be simpler to introduce dimers as a new type of layer. They can be included in our calculations as follows: First, we consider the simplest form of the calculation where we only have dimers. The dimer form factor is related to the monolayer form factor, $F_{\rm d}(h,k,l) = F_{\rm m}(h,k,l)(1 + e^{-2\pi l})$, which is the plane wave from a monolayer at the origin plus an identical layer translated by~$\vec{c}$. Given the doubling of the unit cell, the $z$ component of $R_{ij}$ must be doubled to $2\vec{c}$ and the number of layers is divided by 2 to account for the dimerized layers.

\begin{figure} 
\begin{center}

\includegraphics[width=.95\columnwidth]{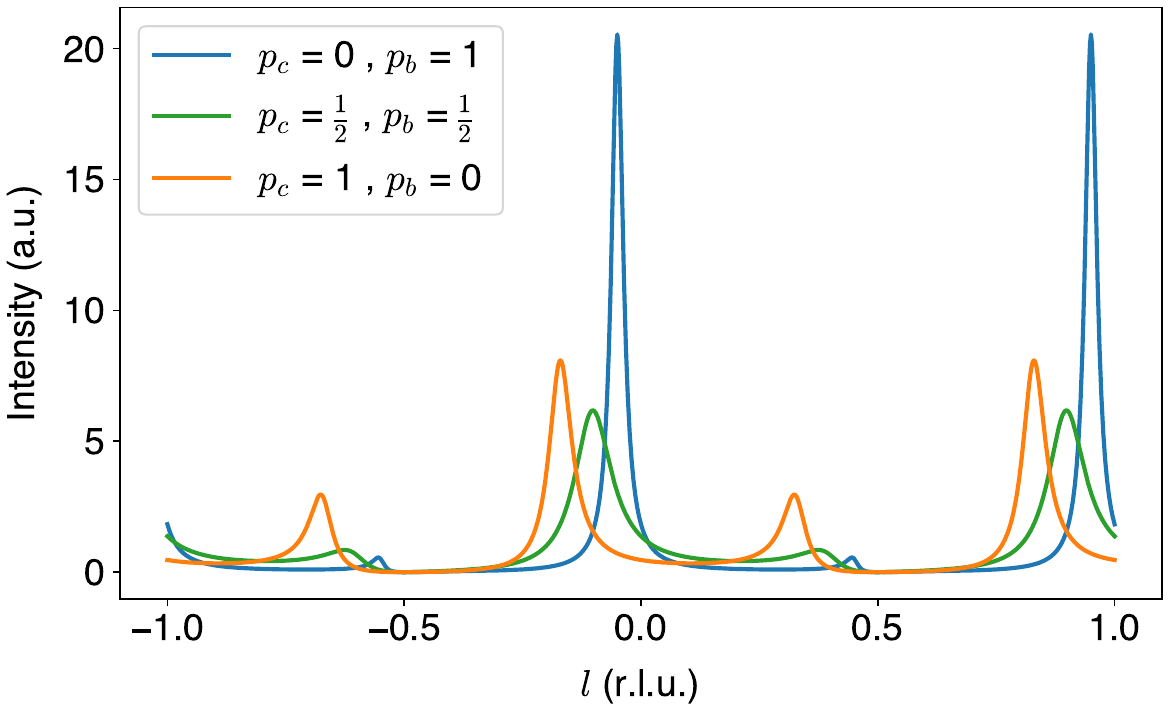} 
\caption{XRD intensity along $l$ for ($h = \frac{1}{13}, k = \frac{3}{13}$) in a fully dimerized state. The intensities for three different stacking arrangements,  from dominant $T_c$, mixed $T_c$ and $T_b$ to dominant $T_b$~stacking, are shown. $p_{c1} = p_{b1} = 1$ for all cases.}

\label{dimer_stacking}

\end{center}
\end{figure}

Consequently, the periodicity of the main peak doubles from $l = \frac13$ r.l.u. in the monolayer calculation with full $T_c$ stacking to $l = \frac16$ r.l.u. in the fully dimerized state, as seen in Fig.~\ref{dimer_stacking}. Additionally, dimer peaks at $\frac12$ r.l.u. away from lattice peaks are also found---a fingerprint of the CCDW state (see End Matter). The stacking configuration with fully disordered $T_c$~stacking is the one proposed in Ref.~\cite{Stahl2020} and produces a periodicity of $6$. However, in experiments the r.l.u. position of the main CDW peak in the CCDW state has been seen to be closer to $l = \frac{1}{5}$ r.l.u. (see~Tab.~1 of the main text), indicating a more complicated stacking structure.
\FloatBarrier 

\subsection{Monolayer-Dimer Coexistence XRD Intensity Simulation}

Simulating a 1\emph{T}-TaS\textsubscript{2} system with a probabilistic mix of monolayers and dimers requires a combination of the calculations mentioned above, treating dimers and monolayers as distinct layer types. For this calculation, we tune the number of dimers relative to the number of monolayers continuously. This introduces a new probability $p_a$ of two $T_a$ stacked dimerized layers above any randomly chosen layer. $p_a$ is then the fraction of layers that are dimerized, in which case we define the dimer as one layer. Therefore, $p_a = \frac{N_{\rm m}}{N_{\rm d}/2 + N_{\rm m}}$, where $N_{\rm m}$ and~$N_{\rm d}$ is the number of monolayers and dimers, respectively. We can define a $2 \times 2$~monolayer-dimer stacking matrix $A^{\rm dm}$ starting with

\begin{equation}
    P^{\rm dm} = 
    \begin{bmatrix}
        p_{\rm mm} & p_{\rm md} \\
        p_{\rm dm} & p_{\rm dd} \\
    \end{bmatrix}.
\end{equation}

\noindent The first column of $P^{\rm dm}$ represents the overall probability to transition from a monolayer to a monolayer/dimer,  whereas the second column represents the probability to transition from a dimer to a monolayer/dimer. From our definition of $p_a$, we can write the elements of this matrix as 
\begin{equation}
p_{\rm mm} = p_{\rm md} = (1-p_a), \hspace{1 cm} p_{\rm dm} = p_{\rm dd}  = p_a. 
\end{equation}

\noindent This would be our matrix if the monolayers and dimers only had one stacking vector, but with 13 possible stacking vectors for each layer type, we must define our overall stacking probability matrix as

\begin{equation}
    A^{\rm dm} =  P^{\rm dm} \otimes A_{ij}^{1},
\end{equation}

\noindent where $A_{ij}^{1}$ is the $13 \times 13$ stacking probability matrix from Eq.~\eqref{stacking probability matrix}. Here, the $26 \times 26$~matrix $A_{ij}^{\rm dm}$ is now formed by four $13 \times 13$ identical NN blocks weighted by the associated monolayer-dimer stacking probability. The columns of this matrix are properly normalized by construction. Additionally, the stacking vector matrix can be written as

\begin{equation}
\begin{aligned}
    R^{\rm dm} &= R_{\parallel}^{\rm dm} \otimes J_{13} + J_{2} \otimes R_{\perp}^{(1)}, \\
    R_{\parallel}^{\rm dm} &= \vec{c}
    \begin{bmatrix}
        1 & 2 \\
        1 & 2 \\
    \end{bmatrix}
\end{aligned}
\end{equation}

\noindent where $J_n$ is the $n \times n$ matrix with all ones and $R_\perp^{(1)}$ is the in-plane stacking matrix from before. The matrix~$R_{\parallel}^{\rm dm}$ represents the magnitude of out-of-plane translation when transitioning between monolayers and dimers. The last complication is that, \emph{a priori}, the probability for a monolayer is different from that for a dimer. This can be accounted for by setting the first 13 elements of the vector $\vec{g}$ to $\frac{1-p_a}{13}$ and the second 13 elements to $\frac{p_a}{13}$, or identically by solving for $\vec{g}$ as the eigenvector with eigenvalue 1. With these definitions of the stacking matrices, the stacking probability matrix $T_{ij}$ can be built and the out-of-plane intensity computed with Eq.~\eqref{Hendriks-Teller Result}. This has been carried out for $T_{c}$ stacking for dimer probabilities ranging from 0 to~1, as seen in Fig. 2 of the main text. The main effect of $p_a$ is that the main CDW peak moves from a periodicity of 3 to 6 continuously and for $p_a$ values larger than $\approx0.5$, a half-ordered peak appears that grows in intensity. $p_a$ is also linked to the correlation length, which is maximized for $p_a = 1$. By only considering the probability $p_a$, the observed periodicity of 5 in \cite{Laulhe2015, Stahl2020} is associated with  $p_a \approx \frac23$. 

\subsection{Dimer Contraction}

To account for the parameter $\epsilon$, defined as the fraction of the $c$ lattice constant that a dimer contracts in the CCDW phase (see End Matter), the dimer structure factor, out-of-plane translation vector, and average out-of-plane lattice constant must be adapted. A parameter~ $c_{\epsilon}$ can be defined that represents the average spacing between layers in an infinite stack of layers and calculated as

\begin{equation}
c_{\epsilon} = c\left( 1 - \frac{p_a \epsilon}{2} \right).
\end{equation}

\noindent Note that if $p_a = 0$, there is no change, and if $p_a = 1$, the average lattice constant is $c(1-\epsilon/2)$. This is the expected result given that half of the out-of-plane layer displacements will be modified by $\epsilon$ and the others will remain $c$. Additionally, the dimer layer form factor must be changed to account for the top layer contracting by $c\epsilon$ downwards, defined as:

\begin{equation}
F_{\rm d}^\epsilon(h,k,l) = F_{\rm m}(h,k,l)\left(1+e^{-2\pi (1-\epsilon) \frac{c}{c_{\epsilon}}}\right).
\end{equation}

\noindent The parallel transition matrix between dimers and monolayers must be changed to the following,

\begin{equation}
R_{\parallel \epsilon}^{\rm dm} = 
    \vec{c}
    \begin{bmatrix}
        1 & 2-\epsilon \\
        1 & 2-\epsilon \\
    \end{bmatrix}.
\end{equation}

\noindent These definitions ensure consistent $l$ locations by constraining that lattice peaks remain at integer $l$-values at every $p_1$, as seen in Fig. 7 of the End Matter.

\subsection{$T_{b}$ Stacking}

\begin{figure*}[!t]
	\includegraphics[width=.65\textwidth]{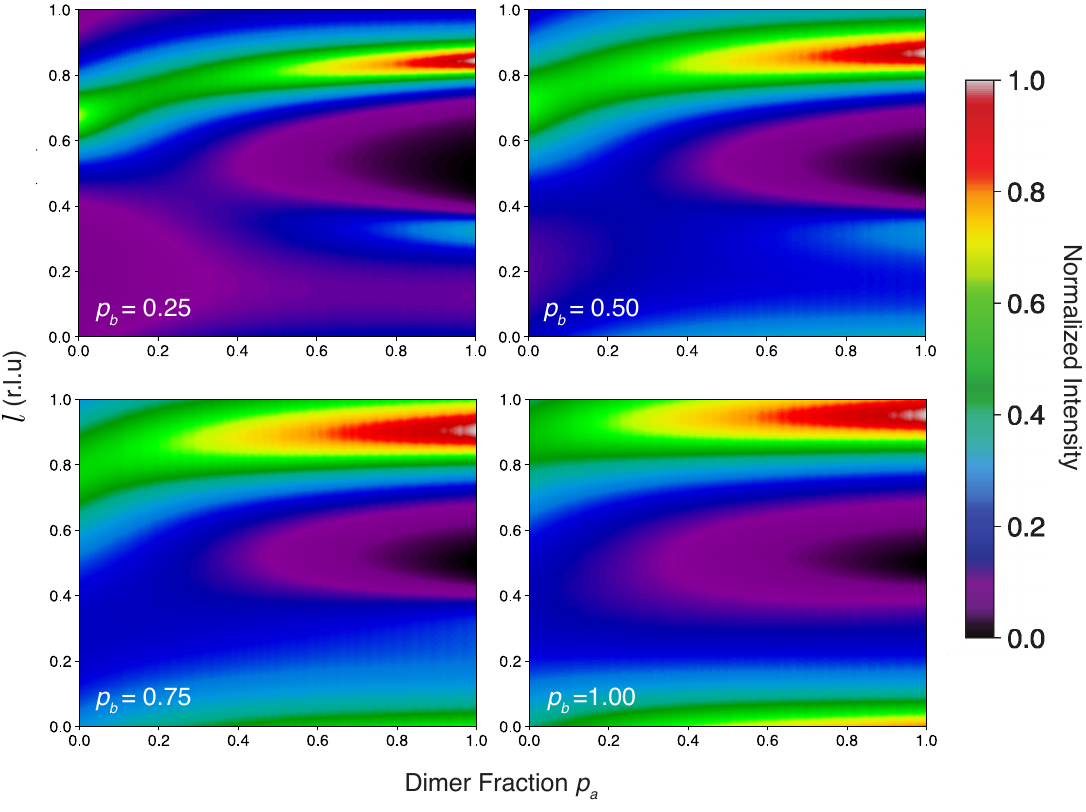}
	\caption{\label{Tb} XRD diffraction intensities  at ($h$ = $\frac{1}{13}$, $k$ = $\frac{3}{13}$, $l$) for various degrees of $T_{b}$ stacking.}
	
\end{figure*}

CDW intensities at ($h$ = $\frac{1}{13}$, $k$ = $\frac{3}{13}$) were simulated by varying the stacking probabilities $p_{a}$, $p_{b}$, and $p_{c}$ with the HT method. This produced a three-dimensional voxel with the axes of $l$ (r.l.u.), $p_{a}$, and $p_{b,c}$ where the latter shows a continuous progression from $T_{c}$ to $T_{b}$~stacking. Figure~2a of the main text shows the 2D slice with $p_{c} = 1$ for various degrees of dimerization~$p_{a}$. Figure~\ref{Tb} shows 2D slices for various degrees of $T_{b}$ stacking ($p_{c} \neq$ 1). As the degree of $T_{b}$ stacking increases for a fully dimerized system, the main CDW peak moves from $l = \frac{1}{6}$~r.l.u. towards integer~$l$, whereas the half-ordered CDW peak disappears. While $T_{b}$ stacking has been observed in surface-sensitive experiments \cite{Butler2020, Wang2023}, our HT~simulations for $T_{b}$ stacking do not align with the experimental XRD data~\cite{Stahl2020}. Based on our simulations, bulk 1\emph{T}-TaS$_{2}$ in the low-temperature CCDW state is likely to contain less than 5\% $T_{b}$ stacking as $p_{b} < 0.05$ makes a negligible difference in the diffraction peak positions for $N$ = 100 layers.

\section{Monte Carlo Simulations}

This section is dedicated to evaluating Eq.~\eqref{Seq2} with a numerical approach. We note that Eq.~\eqref{Seq2} is an expectation value over a statistical system. Drawing from statistical mechanics, $P(x)$ will be a sharply peaked function which can be defined as

\begin{equation} \label{partition function}
P(x) = \frac{e^{- \frac{E(x)}{k_{\rm B} T}}}{Z} \hspace{1.5 cm} Z = \sum_{x \in X} e^{- \frac{E(x)}{k_{\rm B} T}},
\end{equation}

\noindent where $E(x)$ is the total energy of the configuration $x$, $T$~the temperature and $k_{\rm B}$ the Boltzmann constant. For systems with reasonably large $N$ and small frustration, $P(x)$ is a sharply peaked function around a restricted number of ground states. Thus, only a small number of configurations compared to the total number will contribute to the expectation value. An efficient computation method to reduce the number of terms in Eq.~\eqref{Seq2} is to randomly sample over the configurations which have a large $P(x)$, known as the MC method~\cite{Frenkel2017}. One way to find different configurations with large $P(x)$ includes building layers of the material from the bottom up with certain probabilistic rules that encodes the interlayer interactions. Given a configuration of size $n$, called~$x_n$, a layer of type $i$ is added with a Markov chain probability $P(i | x_n)$. After adding it to the stack the process is repeated with~$x_{n+1}$ until the desired number of layers is reached. With a correctly chosen function $P(i | x_n)$, the configurations that are generated will be those that have a large $P(x)$ and contribute most to Eq.~\eqref{Seq2}. In this way, the intensities are not weighted explicitly to execute Eq.~\eqref{Seq2}, but only configurations with large weights are likely to appear.

\subsection{Application to 1\emph{T}-TaS$_{2}$}

\begin{figure*}[!ht] 
\begin{center}

\includegraphics[width=.75\textwidth]{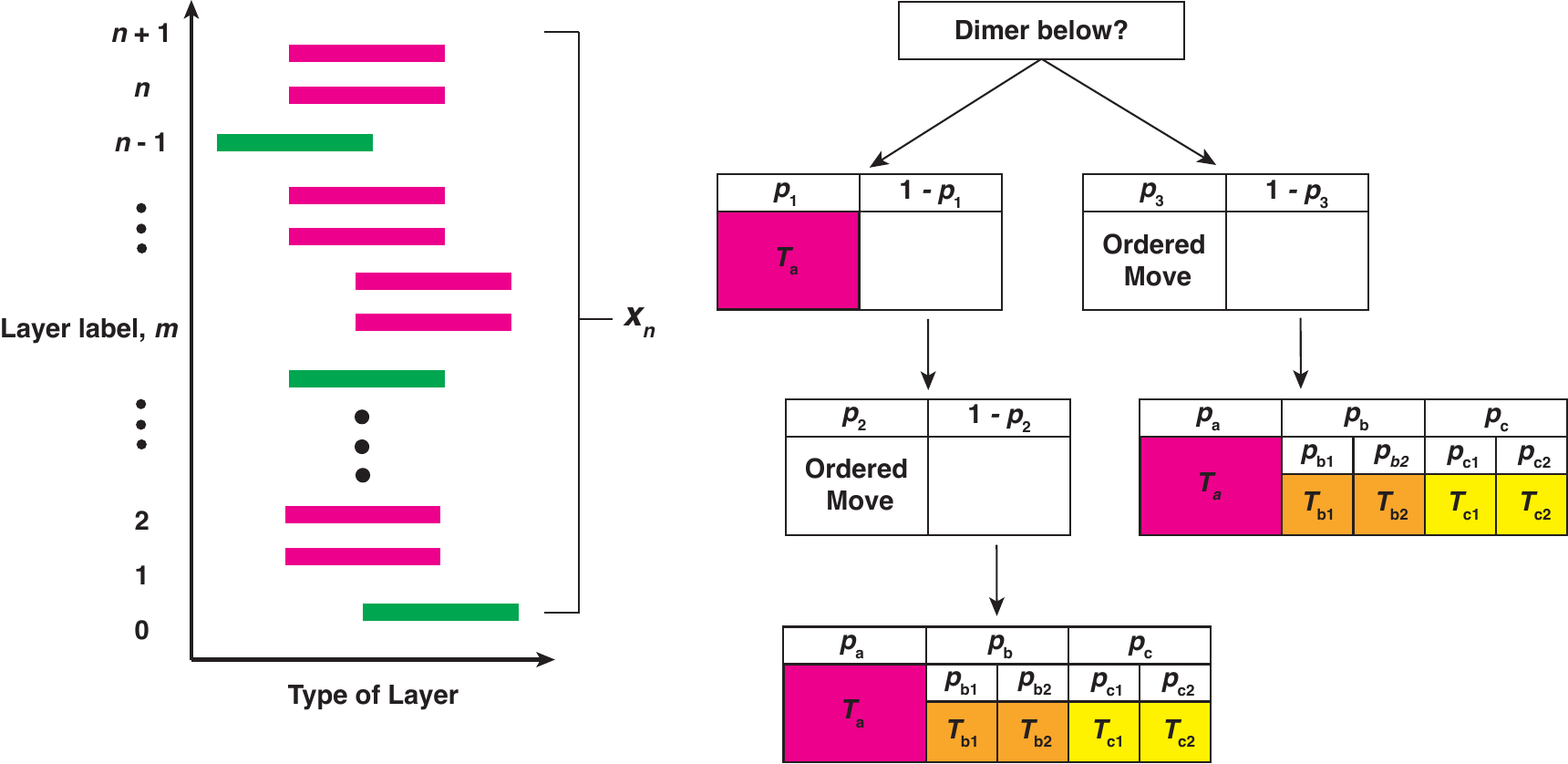} 
\caption{\label{flowchart}Schematic of the $(n+1)^\text{th}$-layer generation on the left by using the flowchart on the right. Monolayers and dimers are colored as green and red, respectively. The terminating boxes of the flowchart indicate the chosen stacking operation. Boxes split into multiple columns representing a stochastic choice based on the labeled probabilities.}

\label{flowchart_MC}

\end{center}
\end{figure*}

For 1\emph{T}-TaS\textsubscript{2}, one way to stochastically generate configurations layer-by-layer is shown in Fig. \ref{flowchart_MC}. A specific configuration of $n$ layers is shown with dimers and monolayers. To stochastically generate the $(n+1)^\text{th}$ layer, the flowchart in Fig.~\ref{flowchart_MC} is algorithmically executed. First, whether or not the top two layers are a dimer is checked. If not, a dimer ($T_a$ stacking) is created with the probability $p_1$ (equivalent to $p_a$ in the earlier sections and the main text). If this does not happen or if there is a dimer below, an ordered move with a probability $p_3/p_2$ is carried out that evaluates a definite function $f(x_n)$ to generate one of the $13$ possible stacking vectors. If an ordered move is not carried out, one of the five stacking groups from Fig. \ref{SFig1} is chosen at random according to the given NN stacking probabilities. This strategy is just an algorithmic representation of a complicated MC chain which encodes the dimer statistics, NN symmetry equivalent translations, and NNN ordering statistics. The probabilities included in this flowchart are representative of the energy of layer-layer interactions. The probabilities $p_2$ and $p_3$ control ordered movement and are in general different because having a dimer below a layer effectively increases the NNN distance and the strength or probability of the ordering behavior. From this physical logic, we would expect that $p_2 \gg p_3$. The probabilities $p_2$~and~$p_3$ function to switch on/off beyond NN interactions. If the interaction energy is large beyond~NN, the degeneracy of the five stacking groups is broken and an ordered structure emerges as the ground state. Twelve structures exist for the 1\emph{T} polymorph of transition metal dichalcogenides that form polar star structures \cite{Walker1983}, which each correspond to specific reciprocal space peak positions. These twelve ordered movements represent the options for ordered movements in the Markov chain. For example, two helical structures are possible that create an in-plane spiral-staircase structure producing a repeating triangular structure with scattering peaks at $l =\pm \frac13$ (see Fig. 6b of the End Matter). A zig-zag structure can exist with peaks at $l = \frac{1}{13}, \frac{3}{13}, \frac{9}{13}, \frac{5}{26}, \frac{15}{26},$ and $\frac{19}{26}$. Each of these ordered stacking structures have a different unit cell and, therefore, correspond to a different point group. This modifies the symmetry of the peak positions in the reciprocal space. With disorder, the peak positions are modified but the symmetry remains the same, rendering it relatively simple to narrow down a specific structure based on experimental data. 

To evaluate Eq.~ \eqref{Seq2}, we used the Monte Carlo chain defined with this flowchart to generate configurations of layers of size $N$. In practice, this is done by generating the minimum number of layers to begin at random, usually three, then ``thermalizing" the system by executing the MC chain $N$ + 30 times and discarding the first 30~layers. This approach functions to throw away the randomly chosen initial layers. A computable large number of configurations, roughly 1000, are generated and then Eq.~\eqref{Seq1} is evaluated for each one and averaged over the configurations $N_{\rm avg}$ to yield an approximate answer to Eq.~\eqref{Seq2}. Namely, the following equation is computed

\begin{equation} \label{monte_carlo_intensity}
    I_{N_{\rm avg}}(\vec{Q}) = \frac{1}{N_{\rm avg}}\sum_{i = 1}^{N_{avg}} I_{x_i}(\vec{Q}),
\end{equation}

\noindent where $x_i$ is the $i^{\rm th}$ configuration that was generated with the repeated stochastic layer approach. A comparison between the analytical HT solution and the statistical evaluation is shown in Fig. 5 of the End Matter.

The Markov chain method presented here approaches the analytical solution for simple NN interactions. The stochastic method is useful for its simplicity in adding beyond NN~interactions, because the Markov chain matrix does not need to be computed---a flowchart is sufficient to generate new layers with any number of NN interactions. Additionally, the analytical method can only be used exactly with a commensurate structure, which reduces the set of stacking vectors to a discrete subset. For an incommensurate in-plane structure, this simplification is not possible. For the nearly-commensurate and commensurate states of 1\textit{T}-TaS$_2$, the analytical approach is still valid given that the in-plane structure involves commensurate domains separated by domain walls. The MC~method was used to generate the ten 20-layer stacking configurations that are detailed in Tab.~S1, where the probability of dimerization was set to $p_{1} = p_a = \frac{2}{3}$ and $p_{c} = p_{c1}=1$ for \emph{T$_{c}$}~stacking of a single chirality.

\begin{table*}[!ht]
\centering
\begin{tabular}{c|cccccccccccccccccccc}
\toprule
\textbf{Configuration} & \multicolumn{20}{c}{\textbf{Stacking Order}} \\
\midrule
1 & ~0~ & ~0~ & 11 & ~9~ & ~9~ & ~3~ & ~3~ & 10 & 10 & ~8~ & ~6~ & ~0~ & ~7~ & ~2~ & ~2~ & ~9~ & ~9~ & ~7~ & ~7~ & ~2~ \\
2 & 0 & 7 & 1  & ~1~ & 12 & 6 & 6 & ~0~  & ~0~  & 8 & 8 & 6 & 6 & 0 & 7 & 7 & 1 & 1 & 8 & 8 \\
3 & 0 & 0 & 7  & 7 & 2  & 2 & 10 & 10 & 8  & 8 & 6 & 6 & 0 & 7 & 7 & 1 & 1 & 9 & 9 & 7 \\
4 & 0 & 0 & 11 & 11 & 5 & 12 & 12 & 6  & 0  & 7 & 7 & 5 & 5 & 3 & 3 & 11 & 11 & 5 & 0 & 11 \\
5 & 0 & 0 & 11 & 9  & 9 & 7  & 5  & 5  & 0  & 0 & 8 & 8 & 6 & 6 & 1 & 1 & 12 & 7 & 7 & 2 \\
6 & 0 & 0 & 8  & 8  & 6 & 6  & 1  & 1  & 12 & 12 & 6 & 6 & 0 & 0 & 8 & 8 & 3 & 3 & 11 & 11 \\
7 & 0 & 0 & 7  & 2  & 9 & 3  & 3  & 1  & 9  & 9  & 7 & 7 & 5 & 5 & 3 & 3 & 1 & 1 & 8  & 8  \\
8 & 0 & 11 & 11 & 5  & 5 & 3  & 3  & 11 & 11 & 6  & 6 & 0 & 0 & 11 & 11 & 6 & 6 & 0 & 0 & 11 \\
9 & 0 & 7  & 7  & 5  & 5 & 3  & 3  & 11 & 11 & 5  & 5 & 12 & 12 & 6 & 6 & 1 & 1 & 8  & 8  & 3 \\
10 & 0 & 7  & 7  & 5  & 12 & 7  & 7  & 2  & 9  & 9  & 7 & 5 & 3 & 11 & 11 & 6 & 0 & 11 & 9  & 9  \\
\bottomrule
\end{tabular}
\caption{Ten 20-layer stacking configurations generated from MC simulations where $p_{1} = \frac{2}{3}$ and the stacking between monolayers and dimers is of $T_{c}$ type. The subsequent stacking of polaron stars is referenced to the center Ta atom 0 of the first polaron star. For example, in configuration~1, the fifth layer is labeled 9 and means that the Ta atom 9 is directly above the center Ta atom of the first layer.}
\end{table*}

\subsection{Electronic Structure}

\begin{figure*}[!ht]
	\includegraphics[width=.9\textwidth]{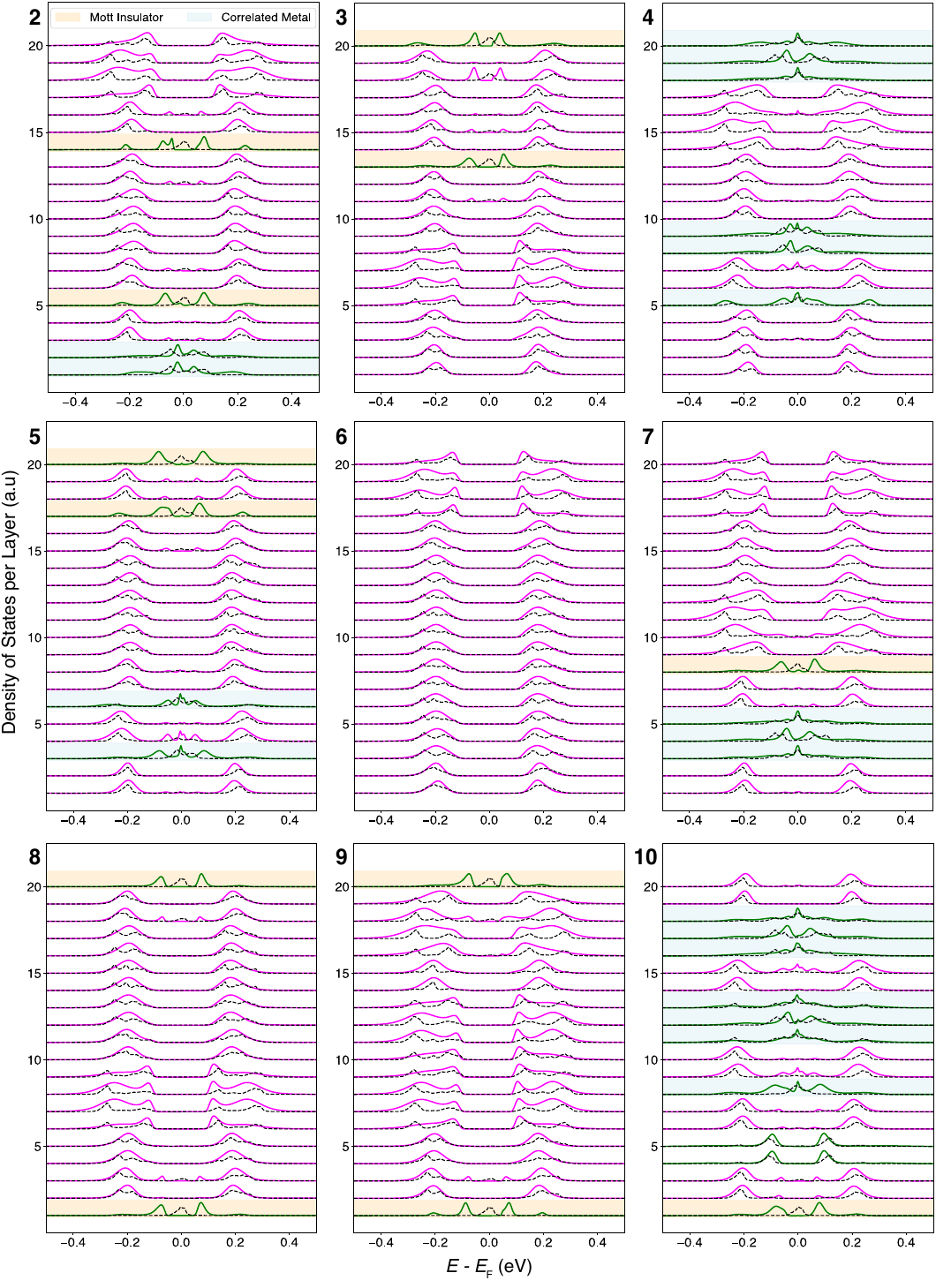}
	\caption{\label{Fig_all_configs}
	Layer-by-layer DOS (black dashed lines) and local spectral functions for configurations~$2-9$ from Tab.~S1. Solid lines correspond to the monolayer (green) and dimer~(pink) local spectral function. Mott insulating and correlated metallic layers are marked by the yellow and green shaded regions, respectively, while the unshaded layers are band insulating.}
\end{figure*}

The layer-by-layer density of states for configuration 1 (see Tab.~S1) is shown in Fig.~3 of the main text, while the remaining configurations are shown in Fig. \ref{Fig_all_configs}. While it is computationally fast to generate large-layer systems with MC, structures confined to \mbox{20-layers} were chosen to compute the density of states (DOS) with dynamical mean-field theory in a reasonable time frame. For example, Configuration 8 in Fig. \ref{Fig_all_configs} is a fully dimerized system with \emph{T$_{c}$} stacking that would result in CCDW diffraction peaks at $l = \frac{1}{6}$ r.l.u. 

It is interesting to note that while all dimerized layers are band insulators, monolayers can be Mott insulating, band insulating, or metallic. Based on our sample size, an isolated monolayer is more likely to be Mott insulating while consecutive monolayers tend to be metallic. There are certain exceptions---for example, layers 4~and~5 in configuration 10 are consecutive monolayers---but turn out to be band insulating with a smaller gap than the dimerized band insulators. In this specific configuration, the two monolayers are squeezed by bilayers that are aligned vertically, \textit{i.e.} the central Ta atoms of the bilayer above and below the monolayers are aligned. In the same configuration, we also see a Mott insulating monolayer (layer 1) and a metallic monolayer (layer~8) with noticeable proximity effects. Therefore, the electronic structure of a monolayer is highly dependent on the local stacking configuration. Within our 10 configuration sample size, we have 20 surface layers of which all monolayers are Mott insulating. This may be due to the fact that proximity effects only come from one direction. Although a larger sample size would be needed to determine whether an isolated monolayer at the surface can be metallic as well, this finding may already explain why surface states as seen by ARPES or STM have only been Mott or band insulating.

\subsection{Ordered Stacking}

The analytical HT approach thus far only includes NN interactions. This is valid in the CCDW state of 1\emph{T}-TaS\textsubscript{2} where dimerization splits NN degenerate stacking vectors. However, for the NCCDW and HCDW states ordering behavior is present, requiring beyond NN interaction terms. Describing greater than first NN~interactions relies on higher than first-order Markov matrices to account for the stacking probabilities. A second-order stacking Markov matrix can be created from $P(x_{n} | x_{n-1}, x_{n-2})$, the probability to stack to a layer type $x_n$ on layers of types $x_{n-1}$ and $x_{n-2}$. If there are $N$~layer types, this represents an $N$ x $(N^2)$ matrix. Likewise, for a $k^{\rm th}$-order Markov process, this matrix is $N \times (N^k)$. An example for the simplest 2-layer case is shown below. The $2\times 4$~stacking probability matrix is populated with $p_x^{yz}$~terms where~$x$ is the layer type to transition to, and $y$~and~$z$ are the first and second NN layers, respectively.

\begin{center}
\begin{tabular}{>{\centering\arraybackslash}  m{0.5cm}  ||>{\centering\arraybackslash}m{1cm}  |>{\centering\arraybackslash}m{1cm} |>{\centering\arraybackslash}m{1cm}   |>{\centering\arraybackslash}m{1cm} } 
    & AA & AB & BA & BB \\ 
  \hline   \hline 

  A & $p_{\rm A}^{\rm AA}$ & $p_{\rm A}^{\rm AB}$ & $p_{\rm A}^{\rm BA}$ & $p_{\rm A}^{\rm BB}$ \\ 
  \hline
  B & $p_{\rm B}^{\rm AA}$ & $p_{\rm B}^{\rm AB}$ & $p_{\rm B}^{\rm BA}$ & $p_{\rm B}^{\rm BB}$ \\  
\end{tabular}
\end{center}

This 2$^\mathrm{nd}$-order Markov matrix can be transformed to an identical first-order one by labeling the rows with a 2-layer index ($x, y$), where $x$ is the generated layer and $y$ is the previous layer. The columns are still labeled with $k,z$. Elements where $y \neq k$ have zero probability. The other matrix elements where $y = k$ are assigned to the value $p_x^{yz}$ from the second-order matrix. With this transformation, our 2$^\mathrm{nd}$--order $2 \times 4$ matrix is represented by the first-order $4 \times 4$ matrix

\begin{center}
\begin{tabular}{>{\centering\arraybackslash}  m{1cm}  ||>{\centering\arraybackslash}m{1cm}  |>{\centering\arraybackslash}m{1cm} |>{\centering\arraybackslash}m{1cm}   |>{\centering\arraybackslash}m{1cm} } 
   
    & AA & AB & BA & BB \\ 
  \hline  \hline
  AA & $p_{\rm A}^{\rm AA}$ & $p_{\rm A}^{\rm AB}$ & 0 & 0 \\ 
  \hline
  AB & 0 & 0 & $p_{\rm A}^{\rm BA}$ & $p_{\rm A}^{\rm BB}$ \\  
  \hline
  BA & $p_{\rm B}^{\rm AA}$ & $p_{B}^{\rm AB}$ & 0 & 0 \\  
  \hline
  BB & 0 & 0 & $p_{\rm B}^{\rm BA}$ & $p_{\rm B}^{\rm BB}$ \\  
\end{tabular}
\end{center}

This can also be done for the general case with $N$~layer types and a $k^{\rm th}$-order Markov process to yield a \mbox{$(N^k) \times (N^k)$} transition matrix by relabeling the rows with multi-layer indices and setting the impossible elements to zero. For the HT model of 1\emph{T}-TaS\textsubscript{2}, the $13 \times 13$ nearest  neighbor $A_{ij}^{(1)}$ transition matrix from Eq.~\eqref{stacking probability matrix} becomes $A_{ij}^{(2)}$, a $169 \times 169$~matrix with a similar transformation. 

In principle, it would also be possible to parameterize the second and third NN interactions with different probabilities that split the degenerate stacking vectors in Fig.~\ref{SFig1}. However, this is complicated and increases the number of independent variables. For this reason, the same approach is carried out with MC instead, which introduces beyond first NN coupling by perfect order where the stacking vector follows a predictable behavior such as a triangular structure or a constant stacking vector. This is done by creating a $13 \times (13^k)$ matrix $B_{ij}^{(2)}$ where each column contains a $1$ for the ordered move. Converted to a $(13^k) \times (13^k)$ matrix $B_{ij}^{(1)}$, the following matrix is formed
\begin{equation}
    C = p_2 B^{(1)} + (1-p_2) A^{(2)}.
\end{equation}

\noindent With this definition, ordered movement takes place with probability $p_2$, just as for the Markov chain represented by the flowchart in Fig. \ref{flowchart_MC}. By constructing stochastic matrices $B^{(1)}$ and $A^{(2)}$, this can be inputted into Eq.~\eqref{Hendriks-Teller Result} to calculate the intensity of an ordered stacking structure. The in-plane structure factor $\vec{F}$ is calculated with the usual method and the initial layer distribution. The vector $\vec{g}$ is found as the stationary eigenvalue of $C$. This method is used to analyze the structure of the NCCDW and HCDW states in the End Matter, where two ordered stacking structures are shown in Fig. 6.

\end{document}